\documentclass[twocolumn,final]{aa}
\usepackage[utf8]{inputenc} 
\usepackage{multirow}

\usepackage{amsmath}
\usepackage{natbib}
\usepackage{graphicx}
\usepackage{subfig}
\usepackage{empheq}
\usepackage[linktocpage=true,colorlinks=true,linkcolor=blue,citecolor=blue,urlcolor=blue]{hyperref}

\def\vec#1{\ensuremath{\mathchoice{\mbox{\boldmath$\displaystyle#1$}}
{\mbox{\boldmath$\textstyle#1$}}
{\mbox{\boldmath$\scriptstyle#1$}}
{\mbox{\boldmath$\scriptscriptstyle#1$}}}}

\newcommand{\Sum}[2]{\sum\limits_{#1}^{#2}}

\begin{document}
\title{Photometry of supernovae in an image series : methods and application to the Supernova Legacy Survey (SNLS)
\thanks{ Based on
  observations obtained with MegaPrime/MegaCam, a joint project of
  CFHT and CEA/DAPNIA, at the Canada-France-Hawaii Telescope (CFHT)
  which is operated by the National Research Council (NRC) of Canada,
  the Institut National des Sciences de l'Univers of the Centre
  National de la Recherche Scientifique (CNRS) of France, and the
  University of Hawaii.} 
}
\titlerunning{Photometry of SNLS supernovae.}
\authorrunning{P.~Astier {\it et al.}}

\author{
P.~Astier\inst{1},
P.~El~Hage\inst{1},
J.~Guy\inst{1},
D.~Hardin\inst{1},
M.~Betoule\inst{1},
S.~Fabbro\inst{2},
N.~Fourmanoit\inst{3},
R.~Pain\inst{1},
N.~Regnault\inst{1}
}

\institute{ Laboratoire de Physique Nucléaire et des Hautes Energies,
  UPMC Univ. Paris 6, UPD Univ. Paris 7, CNRS IN2P3, 4 place Jussieu
  75005, Paris, France \and 
Department of Physics and Astronomy
  University of Victoria Elliott Building 101 3800 Finnerty Road (Ring
  Road) Victoria, BC, CA \and 
Integral Data Center for Astrophysics,
  Department of Astronomy, University of Geneva ch. d'Écogia 16,
  CH-1290 Versoix, Switzerland.  }

\date{Received Apr 8, 2013; accepted Jun 12, 2013}

\abstract {} {We present a technique to measure lightcurves of
  time-variable point sources on a spatially structured background
  from imaging data. The technique was developed to measure light curves of SNLS
  supernovae in order to infer their distances. This photometry
  technique performs simultaneous PSF photometry at the same sky
  position on an image series.} {We describe two implementations of
  the method: one that resamples images before measuring fluxes, and one which does not. In
  both instances, we sketch the key algorithms involved and present
  the validation using semi-artificial sources introduced in real
  images in order to assess the accuracy of the supernova flux
  measurements relative to that of surrounding stars. We describe the methods
  required to anchor these PSF fluxes to calibrated aperture catalogs,
  in order to derive SN magnitudes.}{ We find a marginally
  significant bias of 2 mmag of the after-resampling method, and no bias at
  the mmag accuracy for the non-resampling method. Given surrounding
  star magnitudes, we determine the systematic uncertainty of SN
  magnitudes to be less than 1.5 mmag, which represents about one third of the current
  photometric calibration uncertainty affecting SN measurements.
  The SN photometry delivers several by-products: bright star PSF flux
  measurements which have a repeatability of about 0.6\%, as for aperture measurements; we measure relative astrometric positions with a noise floor
  of 2.4 mas for a single-image
  bright star measurement; we show that in all bands of the MegaCam
  instrument, stars exhibit a profile linearly broadening with flux 
  by about 0.5\% over the whole brightness range.}  {}
\maketitle
\section{Introduction}
\label{sec:introduction}

Measuring light curves of variable stars is nowadays mostly carried
out by performing photometry in an image series. This usually does not
reduce to just gathering measurements obtained independently on each
image, because one can take advantage of at least two specific
features: most of the stars in the image series are not variable, and
the relative positions of astronomical sources are constant in the
image series, with a subset possibly affected by significant proper
motions.

Most of the proposed implementations of light curve measurements are
able to detect flux variations, below the fluctuations due to
atmospheric extinction or instrumental response variation,
that typically affects
a single-image or single-epoch measurement. Relying on the assumption
that stars are on average non-variable allows one to correct for
these noise sources on an image
per image basis. Measurements aimed at detecting micro-lensing, planet
transits, or more generally measuring small luminosity variations, are
commonly characterised by the smallest relative flux variation they
can detect for their brightest stars. Ground-based instruments can
reach the milli-magnitude level \citep[e.g.][]{2007A&A...470.1137M},
while space-based instruments approach $10^{-5}$
\citep[e.g.][]{Jenkins10}.

Our variable-source photometry pipeline aims at measuring light curves
of supernovae (SNe), in order to infer luminosity distances from
those, in the context of ground-based observations. This requires to
derive some apparent luminosity indicator from the data that is both accurate (i.e. precisely calibrated) and limited only by shot noise (because distant supernovae are faint).
Supernova observations could
in principle be directly calibrated to standard stars. This route is
however inefficient because a sizable fraction of observing nights is
non-photometric, i.e. the temporal or spatial variability of
atmospheric extinction is too large to allow one to reliably assume
that science and calibration targets were observed under sufficiently
similar conditions. Hence, most ground-based supernova surveys
calibrate their supernovae via a two-step process: measuring the ratio
of SN flux to some surrounding stars (step 1), and measuring the fluxes of
these surrounding stars with respect to some standards, usually in a
subset of images (step 2). As these standards are usually secondary standards,
commonly the Landolt catalog \citep{Landolt92} or the Smith catalog,
\citep{Smith02}, the ``relay stars'' in SN fields are called
tertiary stars. This two-step process allows one to rescue 
non-photometric observations of the SN, again under the
assumption that tertiary stars are on average non-variable.
Note, however, that variable stars can be detected and ignored. The key performances
of an SN light curve pipeline is no longer the smallest detectable
luminosity variation, but rather the statistical efficiency\footnote{In statistics, the efficiency of an estimator is defined as the ratio of the 
minimum variance bound (from the Cram\'er-Rao inequality) to its actual variance
(see e.g. \citealt{Kendall61}, vol. 2 \S 18.15).}  of the
supernova measurements and the fidelity (on average) of the ratio of
supernova flux to that of neighbouring stars. These two qualities
are usually called precision and accuracy respectively.

In this paper, we will concentrate on the first measurement step,
i.e. measuring the ratio of SN fluxes to that of tertiary stars, in
the framework of the SuperNova Legacy Survey (SNLS, described in \S
\ref{sec:SNLS-survey}). We will discuss as well 
the comparison of obtained instrumental magnitudes to
calibrated magnitudes obtained at step 2, which turns out
to be more subtle than one might naively think.
We will not discuss here the derivation of 
magnitudes of tertiary stars on a photometric system accurately related
to physical fluxes, but rather point 
interested readers to \cite{Betoule13} (and references therein) for
the data set discussed in this paper, and to \cite{Ivezic07,Tucker06} (and references therein) for a
parallel work on the SDSS SN survey, with some updates in \cite{Betoule13}. 

Regarding the SN-to-tertiary stars
measurement, our approach consists in fitting a time-variable point source
on top of a time-independent galaxy image to the image series. We propose
two incarnations of the procedure, one which requires resampling the images
prior to the fit, and a second one which does not. 
The former was used for past SNLS publications \citep{Astier06,Guy10},
and the latter is very similar to the ``Scene modeling''
described in \cite{Holtzman08} that was developed for the SDSS SN survey.

SN cosmology now requires accurate SN fluxes: with the current sample
of $\sim$ 500 well-measured SNe distances, photometric calibration
uncertainties, typically better than 0.01~mag, contribute as much as
random errors (shot noise and SN variability) to the cosmological
parameters uncertainties \citep{Conley11,Sullivan11}. Since biases of
SN flux measurements relative to field stars contribute to the overall
cosmology uncertainty budget in the same way as photometric calibration
uncertainties, SN photometry is now to be challenged at the few mmag
level. In this work, we report on tests and effects at this level 
of accuracy, or better. Random errors affecting SN measurements are
much larger, but average out. 

The plan of this paper goes as follows: we first briefly describe the
SNLS survey (\S~\ref{sec:SNLS-survey}) and then sketch
(\S~\ref{sec:pre-reduction}) the pre-reduction steps applied to images
prior to SN photometry. We compare our approach with others that have been used 
in \S~\ref{sec:SN-photometry-techniques}.
We then describe our two implementations of the SN photometry,
the ``resampled simultaneous photometry'' (RSP thereafter, \S~\ref{sec:RSP})
and the ``direct simultaneous photometry''
(DSP thereafter,\S~\ref{sec:DSP}).  For the latter we detail the calculation
of the simultaneous astrometric solution and the influence of
atmospheric refraction.  We then enter into the tests of both methods
using simulations that heavily rely on real images (\S~\ref{sec:simulations}).
How we relate instrumental magnitudes of tertiaries to calibrated magnitudes
of the same stars is described in \S~\ref{sec:calibration-using-field-stars}.
We assess the quality of SN photometry  in \S~\ref{sec:sn-photometry}.
We discuss the variation of PSF size with star brightness in \S~\ref{sec:brighter_fatter}. We briefly sketch the salient technical points of the
implementation in \S~\ref{sec:implemention}, and conclude in \S~\ref{sec:conclusion}.
 
\section{The SNLS survey \label{sec:SNLS-survey}}
We deliver here the minimum information about the survey required for what follows,
see \cite{Astier06} for more details. The SuperNova Legacy Survey
(SNLS) was a two-prong survey: the photometry was acquired within the 
deep survey of the Canada-France Hawaii Telescope Legacy Survey
(CFHTLS\footnote{\href{http://www.cfht.hawaii.edu/CFHTLS}{http://www.cfht.hawaii.edu/CFHTLS}}),
conducted on the CFHT from 2003 to 2008, using the then new 1~deg$^2$
imager MegaCam. SNLS also conducted a spectroscopic survey relying mostly on VLT, Gemini
and Keck that we will not discuss further. 
MegaCam \citep{MegacamPaper} gathers 36
back-illuminated thinned CCDs (E2V CCD42-90) of 2048 $\times$ 4612
pixel$^2$ with a plate scale of 0.185 \arcsec/pixel. This plate scale
delivers images which sample typical PSFs with more than 4 pixels FWHM, 
falling to $\sim 2.5$ for the best image quality. These CCDs
are arranged in 4 rows of 9 chips, each covering 6.3$\times$14.2~arcmin$^2$.
The deep CFHTLS
survey consisted in monitoring 4 to 5 times per lunation in the $griz$
bands, 4 pointings spread in right ascension, as long as they remained
visible. Each visit typically consisted in 5 to 8 consecutive images
with exposure times of a few hundred seconds, and ditherings of at
most 250 pixels in right ascension and 1000 pixels in
declination. Most of the observing nights also have calibration
exposures of Landolt fields, in $ugriz$ bands.  The four science fields,
\citep[see table 1 in][]{Astier06} were selected for their low Galactic extinction, and
hence have a low stellar density. MegaCam observations are grouped
in ``runs'', lasting 14-18 nights in a row, centered on new moon.
The camera is removed from the telescope during bright time. 
MegaCam observations are acquired by the observatory staff
according to observers' requests. 
Depending on band and field, the CFHTLS deep survey has
delivered 500 to 800 individual exposures that are used to measure
lightcurves of supernovae.

All images gathered with MegaCam have very similar orientations
(relative rotations are of the order of 0.2$^\circ$ rms, but much less within a run), and the $x$
and $y$ coordinates of the CCDs are fairly well aligned with right ascension and
declination. At the beginning of the survey, the image quality was
typically 20\% worse in the corners of the focal plane than in the
center. This improved to $\sim$10\% after the flip of the L3 lens of
the image corrector in Dec 2004. In July 2007, the $i$ filter was
accidentally broken, and a replacement filter was procured within 3
months, slightly different from the original, which we call $i2$ 
in what follows. No SN event has data in both filters.

\section{Pre-reduction and PSF modeling}
\label{sec:pre-reduction}
MegaCam images are processed at CFHT before release using
the Elixir pipeline \citep{Magnier04}. This set of tools assembles
flat-field frames from twilight images from a whole MegaCam run
and applies those consistently to all exposures. It also extracts
fringe patterns from all science images in $i$ and $z$ bands and
subtracts those. Images are delivered with an astrometry to $\sim$1\arcsec.

Some processing of the images is required before they can enter
the SN photometry pipeline. We typically need:
\begin{itemize}
\item some estimates of data quality, e.g. image quality (IQ), objects counts, 
a preliminary estimate of the photometric zero point in order to assess
the atmospheric extinction;
\item a map of pixel weights initialised from inverse sky variance and flat-field frames. This map also identifies the pixels
to be ignored for measurements, typically from CCD defects, saturation, 
cosmic rays and satellite trails;
\item a World Coordinate System (WCS) for each image, obtained from 
matching the image catalog to that of a deep image stack itself anchored
to the USNO catalog to set scale and orientation. We only rely on relative
positions from these WCSs in order to match catalogs from different 
images of the same field;
\item a PSF model for each image allowing for spatial variations.
\end{itemize}

The reductions described in this paragraph are carried out
independently for each CCD (2048 $\times$ 4612 pixels, 6.2$\times$14.2
arcminutes). 

Some images were acquired at CFHT for the survey, but are not part of
the CFHTLS data sample because of their poor quality. We anyway
collect those poor quality images and apply two quality cuts: we reject images with IQ$>$3.5
(defined in Eq.~\ref{eq:iq-definition} below, and
the median IQ is around 2), or with an atmospheric extinction above
$\sim$ 2 magnitudes from the average. In both instances, these images
do not convey a significant amount of information for SN lightcurves.

\subsection{Image catalog and sky background}
\label{sec:image-catalog}
We use SExtractor \citep{Sex} to build a first image catalog, and
obtain a ``segmentation map'' (i.e. a map of pixels attributed to
detected objects). We enlarge the footprint of objects by 5 pixels,
and then compute a sky background map using only unmasked pixels,
using an algorithm similar to the SExtractor one. We then
subtract this smooth background component, and compute the
Gaussian-weighted first and second moments of all detections. The
second moments of the Gaussian weighting function are iteratively adjusted 
to the ones of each object, i.e. the matrix of weighted second moments should
satisfy:
\begin{align}
{\vec M_g} & = 2 \frac{\sum_{pixels} ({\vec x_i}-{\vec x_c}) ({\vec x_i}-{\vec x_c})^T W_g({\vec x_i}) I_i}{\sum_{pixels} W_g({\vec x_i}) I_i}
\label{eq:gaussian-moments} \\
W_g({\vec x_i}) & \equiv \exp \left[ -\frac{1}{2} ({\vec x_i}-{\vec x_c})^T{\vec M_g}^{-1}({\vec x_i}-{\vec x_c}) \right ] \nonumber
\end{align}
where ${\vec x_i}$ are pixel coordinates, ${\vec x_c}$ the Gaussian weighted
centroid obtained similarly, and $I_i$ is the (sky subtracted) image
value at pixel $i$. This iterative adjustment of second moments mostly
fails on extremely sharp detections typically due to image defects or
cosmic rays. The algorithm often diverges on blended objects, which
then do not get second moments measurements. Equation \ref{eq:gaussian-moments}
is the normal equation for second moments of a least squares fit of
a 2D-Gaussian to the image, {\it assuming a stationary noise} (i.e. position-independent). Ignoring 
the contribution of the object to pixel variance makes the relative
weights of pixels independent of the star flux, and hence
ensures that the inadequacy of the Gaussian PSF to describe
the actual star shapes does
not cause a flux-dependent shift of these Gaussian second moments.  
We have checked on simulated images with a non-Gaussian
PSF that this size estimator is independent of brightness at the 
$10^{-4}$ level.

\begin{figure}[h]
\centering
\includegraphics[width=\linewidth]{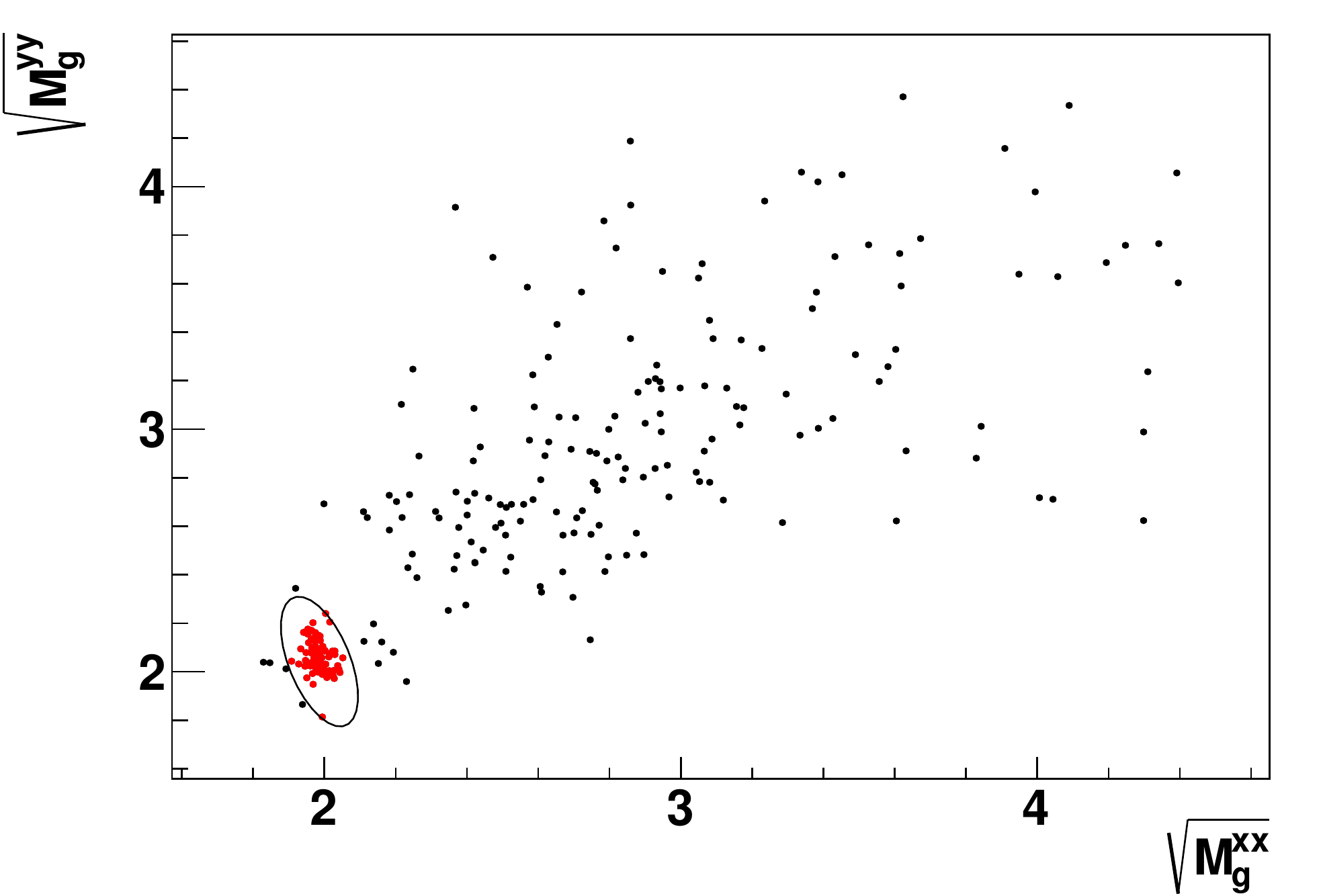}
\caption{Gaussian-weighted second moments from a single typical image,
with the found star clump and the star selection (red points 
within the ellipse). 
\label{fig:second_moments}
}
\end{figure}

The Gaussian-weighted second moments of stars tend to cluster in
the $(M_g^{xx},M_g^{yy})$ plane. The shape of the star clump (due to the
variation of PSF across the CCD) is modeled as a 2-D Gaussian
distribution and stars are selected within a 5-$\sigma$ ellipse, as
shown in figure \ref{fig:second_moments}. 
We compute the average second moment matrix of found stars $\bar{M}$
and define the image quality as:
\begin{equation}
\sigma_{IQ}\equiv \sqrt[4]{\det (\bar{M})} \label{eq:iq-definition}
\end{equation}
$\sigma_{IQ}$ hence refers to a Gaussian r.m.s rather than FWHM, and is expressed in
pixel units. A 0.8\arcsec~FWHM seeing, typical for the CFHTLS
observations, translates to $\sigma_{IQ}\sim 2$ with our definition. We define
a set of circular apertures in units of $\sigma_{IQ}$, and measure the fluxes
of all detections in these apertures, keeping track of bad or
saturated pixels, and pixels attributed to other detections. These
aperture catalogs constitute the basic bricks of the tertiary catalogs
which are presented in \cite{Betoule13}.  Images in which the star
cluster cannot be found are ignored for further processing. These
failures are usually due to massive extinction or severe guiding
errors. The WCSs are computed from the Gaussian-weighted positions. 

\subsection{PSF modeling \label{sec:PSF-modeling}}
The star catalog from each image is used as input for modeling the PSF.
We roughly follow the strategy of DAOPHOT \citep{DAOPHOT}: modeling an
analytic point spread function (PSF) and complement it with pixelised correction at the same
sampling as the images. The analytic part offers the advantage that
it accurately describes the dependence of the PSF with respect to the
object position within the central pixel, and the non-analytic part
accommodates departures from the analytic shape, such as asymetries and
guiding errors. For the analytic part, we chose an elliptical  Moffat PSF \citep{Moffat69}:
\begin{align}
P(x,y) & = A \left[ 1+ r^2 \right]^{-\beta}, \nonumber \\
  r^2 &\equiv w_{xx}x^2+w_{yy}y^2+2w_{xy}xy \nonumber \\
  A & \equiv \frac{\beta-1}{\pi}\sqrt{ w_{xx} w_{yy} - w_{xy}^2} \nonumber
\end{align}
with $\beta=2.5$.
Both the parameters of the analytic part ($ w_{xx},w_{yy},w_{xy}$)
and the pixels of
the non-analytic part are modeled as linear functions of position
within the CCD.
All stars are used as PSF models (with the standard least squares
pixel weighting, see \S\ref{sec:psf-photometry}), and we robustify the fit by eliminating stars and single
pixels which have a deviant contribution to least squares. The process
is entirely automatic and, as for star finding, failures are commonly due to large
extinctions or severe guiding errors. The outcome of the process 
is a PSF model as a function of position in the CCD, and the flux and
position of stars, with their uncertainties, where the latter only
account for shot noise of sky and objects. Due to their sharing of the same (uncertain)
PSF model, parameters of different stars are correlated, but these
small correlations are ignored in what follows.

\subsection{A few technical points about PSF photometry}
\label{sec:psf-photometry}

We refer to Appendix B of \cite{Guy10} for a discussion of the effects
of position uncertainty on PSF flux estimates, and recall here the
salient points. For a Gaussian PSF, a position error underestimates
the flux by:
\begin{equation}
	\frac{\Delta f}{f} = \frac{1}{4} \frac{\delta x^2 + \delta y^2}{\sigma _{IQ}^2}
	\label{eq:deltaF}
\end{equation}
which is quadratic in the position error. It therefore does not
average out from one measurement to another, and leads to a systematic
bias inherent to PSF photometry. More generally, a PSF flux estimation
on a single image suffers from a bias at low S/N:
\begin{equation}
\mathrm{E}[\widehat{f}] \simeq  f \left\{ 1 - \frac{\mathrm{Var}[\widehat{f}]}{f^2} \right\} \label{eq:bias_var_flux}
\end{equation}
where the approximation obviously breaks down when S/N approaches 1.
Since we have to cope with measurements of SNe at low S/N (we
occasionally deal with $S/N<1$), we have to fit or impose a single
common position on all images. Since we are concerned by the accuracy
of flux ratios, the tertiary stars should also be measured imposing a
common position, so that they are affected by inaccuracies of
coordinate mappings between images in the same way as SNe. 

A least-squares PSF flux estimator reads:
\begin{equation}
\hat{f} = \frac{\sum_i w_i P_i I_i}{\sum_i w_i P_i P_i} \label{eq:psf-flux-estimator}
\end{equation}
where $P$ is the PSF, $I$ is the sky-subtracted image, $w$ denotes the
pixel weights in least squares, and the sums run over pixels.  The
statistically optimal weights read $w_i^{-1}=\mathrm{Var}[I_i] = \mathrm{Var}[sky]+k f P_i$,
where k is the ratio of a pixel content to its shot noise variance,
usually the inverse of the gain. For a faint source, $\hat{f} \propto
\sum_i P_i I_i$ and for a bright source $\hat{f} \propto \sum_i I_i $,
so that the relative weights of image pixels $I_i$ vary with source
brightness. Flux ratios are then accurate only if the PSF model is
faithful. Setting $w_i^{-1} = \mathrm{Var}[sky]$ preserves the statistical
optimality for faint sources and makes flux ratios independent of the
accuracy of the PSF model, at the expense of a suboptimal flux
estimator for brighter sources. Since the flux ratio uncertainty is
dominated by the uncertainty of the fainter source, and we have
several tertiary stars for each SN, we settled for $w_i^{-1} = \mathrm{Var}[sky]$
for {\it both} the photometry of SNe and tertiaries. Note that the
reason for assuming a stationary noise when estimating Gaussian second
moments (Eq. \ref{eq:gaussian-moments}) is essentially the same.

We however use the optimal pixel weights (i.e. account for all noise
sources including the object itself) when modeling the PSF (\S
\ref{sec:PSF-modeling}), in order to obtain a PSF model as faithful as
possible. It is worth stressing that there is a systematic
  difference between using $w_i^{-1}= \mathrm{Var}[sky]+k f P_i$ and $w_i^{-1}=
  \mathrm{Var}[sky] + k I_i$ in expression \ref{eq:psf-flux-estimator},
  although these two expressions should agree on average. With the
  second expression, the flux estimator becomes seriously non linear
  w.r.t pixel values $I_i$ and this leads to unacceptable flux biases,
  analogue to the ones described in \cite{Humphrey09}.  The PSF
modelling yields PSF fluxes of tertiary stars, but those will {\it
  not} be used for comparison with SNe, because they rely on
flux-dependent weights and are not obtained by enforcing a common
position on all images.

\section{Overview of SN photometry techniques}
\label{sec:SN-photometry-techniques}
Photometry of variable sources is required to build light curves.
Supernovae are not just like variable stars, because they usually
appear in galaxies, which constitute a spatially structured background
to the SN light. The classical sky subtraction algorithms assume a
spatially smooth background \citep[see e.g.][and references therein]{Irwin85} and hence cannot be used. As most galaxy subtraction schemes,
we will rely on images of the field without the supernova 
acquired either before or well after the explosion. Note that
as mentioned above, tertiary stars (i.e. stars surrounding the 
supernova measured in the same frames) should be measured as well
and in such a way that flux ratios are as accurate as possible. 

Many approaches have been proposed for this differential photometry
problem:
\begin{enumerate}
\item Measure fluxes in the same aperture on both ``on'' and ``off''
  images and subtract those, after a proper flux scaling
  \citep[e.g.][]{Perlmutter99}. Surrounding stars are measured using
  the same aperture photometry on both sets of images (and these
  fluxes indeed define the flux scales).
\item Register (via resampling) the ``off'' image to align it on the ``on'' image (or
  vice versa), match the PSFs (by application of a convolution kernel to the best-IQ image), flux scale one using stars, subtract
  images, and measure the SN PSF flux on the subtraction
  \citep[e.g.][]{Hamuy94,Schmidt98}. Surrounding stars are measured
  using PSF photometry on un-subtracted frames.
\item From the image series, compute all flux differences using the
  above method. There are N(N-1)/2 such differences and the method is
  called NN2 \citep{NN2}. From this potentially large number of
  differences, one fits for the actual light curve points by least
  squares, imposing a null flux constraint. Surrounding stars are
  directly measured using PSF photometry on unsubtracted frames.
\item Fitting a time-independent pixellised galaxy model and a time-variable point
  source to the image series, forcing the flux to zero in images where
  the SN is ``off''. Surrounding stars are measured using the same
  technique without fitting a galaxy and without ``off'' periods
  \citep{Fabbro01,Astier06,Holtzman08,Guy10}.
\end{enumerate}

The last approach directly fits the model to the whole data set
and might be optimal from a statistical point of view by reaching
the minimum variance bound set 
by the Cram\'er-Rao inequality. The NN2 technique could reach
statistical optimality if it tracked the covariance of 
all flux differences when computing the actual light curve.
The method is however computationally prohibitive when dealing with
several hundred images. The
technique sketched in point 2 above approaches 
statistical optimality because there is not much information
to gain about the galaxy light distribution under the SN from images
with the SN. By not using PSF photometry, the method 1
is suboptimal from a statistical point of view,
but independent of any PSF modeling, at variance with all other
approaches. Method 1 also does not require image resampling, at variance
with methods 2 \& 3. 

This paper will discuss two incarnations of the full model technique
(method 4 above): one method assumes that all images are on the same pixel
grid and hence requires resampled images, and the other 
resamples the model rather than the input data, as originally
proposed in \cite{Holtzman08}. In both instances, we use the
weighting scheme discussed in \S \ref{sec:psf-photometry} which avoids
the shortcomings of inaccurate PSF modeling.

\section{Resampled simultaneous photometry (RSP)}
\label{sec:RSP}
The first step of the resampled simultaneous photometry (RSP)
consists in identifying the best IQ image of the series (``the
reference''), and resample all other images to the same pixel
grid as this image. The needed geometrical mappings are fitted to the catalogs and leave
residuals typically around 0.15 pixel (dominated by shot noise). Note that we also resample the
weight maps, in particular to properly account for pixels with null
weight. Then, discrete convolution kernels are fitted to the aligned
image pairs in order to match the PSF of the reference to all the
other images of the series, using the \cite{Alard98} algorithm. More
precisely, we fit a spatially variable kernel \citep{Alard00}, which
can compensate minor misalignment residuals, and we impose
a position-independent kernel integral. The fit is carried out on stamps
centered on the 150 (non-saturated) objects in the frame with the highest peak flux.
Note that the fitted kernel matches both 
the PSF shape and the flux scale of the involved image pair.

On the set of aligned images, the model for the expected light in
image $i$ at pixel $p$ reads:
\begin{equation}
M_{i,p} = \left \{ \left [ f_i \times \phi_{ref} ({\vec x_p} - {\vec x_{SN}} ) + gal_{ref} \right ] \otimes K_i \right \}_p +s_i
\label{eq:light_model}
\end{equation}
where $f_i$ is the SN flux in image $i$, $gal_{ref}$ is the galaxy pixel map
in the reference image (assumed to be non-variable in time) at the sampling of this reference image, $K_i$ is the convolution kernel to match the reference image PSF $\phi_{ref}$ and flux scale
to the ones of image $i$. $s_i$ is the sky of image $i$.
This model is compared to data using least squares:
\begin{equation}
\chi^2 = \sum_i \sum_p  w_{i,p} ( M_{i,p} - I_{i,p})^2
\end{equation}
where $I_{i,p}$ is the data. The fit parameters are one flux per
``on'' image, a single SN position, the galaxy pixel map, and a sky
level per image, except for one image, because it would be degenerate with a
spatially-constant flux added to the galaxy map\footnote{The model of eq. \ref{eq:light_model} is unchanged if one operates the simultaneous substitutions: $gal_{ref} \leftarrow gal_{ref}+C$ and $s_i \leftarrow s_i - C \int K_i$.}. We only fit a stamp
around the SN, typically of 10 times $\sigma_{IQ}$ on a side. Convolution kernels
are also pixel maps whose size is adjusted to the IQ difference
they are expected to bridge. The galaxy map is fitted up to
the size required by the worst IQ image in the series, typically 50 pixels
on a side. 

When fitting a supernova, we impose $f_i=0$ on images 
acquired before or long after the explosion. When fitting the light curve
of a tertiary star, we impose that the underlying galaxy $gal_{ref}$
is zero, and also allow sky level $s_i$ to vary in all images (as opposed 
to freezing one to zero when fitting the galaxy). 

One might note that the fit assumes pixels to be independent, thus
ignoring the correlations introduced between neighbouring pixels by
resampling, because the latter are not easily tractable.  For linear
least squares, approximating the uncertainties is not a source of
bias, but is suboptimal, as stated by the Gauss-Markov theorem\footnote{The Gauss-Markov theorem 
states that for Gaussian problems, among linear estimators, the one with smallest variance maximises the likelihood.}.
Because the object's position does not enter linearly in the fit, 
the argument does not strictly apply and we
will discuss shortly realistic simulations that may detect
a possible flux bias. Regarding optimality, simplified simulations show that
ignoring correlations due to resampling when measuring PSF fluxes on
resampled images has a negligible effect on the real variance
of the flux estimator, for the typical spatial sampling we are 
considering here. Because resampling introduces mostly positive
correlations between neighbouring
pixels, the flux variance estimated from propagating
the apparent sky variance is usually underestimated.

\section{Direct simultaneous photometry (DSP)}
\label{sec:DSP}
The direct simultaneous photometry (DSP)
aims at avoiding any resampling of the data. 
Since input pixels are then uncorrelated, ignoring correlations is no longer an approximation. Propagation of the shot noise is
tractable, and one saves the computer mass storage corresponding to resampled images.
The method can be used on under-sampled images (fitting an over-sampled
galaxy model). The tests presented in \S~\ref{sec:simulations} are more complete than for the RSP method.

The DSP method requires a PSF model for
each image in the series, and coordinate transformations that map
images one on the other. We have already discussed the production of
the PSF of each image in \S\ref{sec:PSF-modeling}, and we will first
describe how we obtain the necessary coordinate mappings.  We then
discuss atmospheric refraction and star positions because position
variations matter for fluxes in the context of a fit that imposes a common
position on all images.  We eventually describe the fit itself.

\subsection{Relative astrometry of the image series}
\label{sec:astrometric-fit}
Since we are fitting SNe simultaneously to an image series, imposing
a fixed position on the sky,
we have to transform its position (in some frame) into pixel coordinates in
any image of the series. Since we are going to use these transformations
to position a PSF on each image, these transformations should be determined
using coordinates obtained using the same PSF model as the one
that will be used during the fit. We hence adjust transformations using
PSF-fitted coordinates of stars,
which are a by-product of PSF modeling (\S~\ref{sec:PSF-modeling}).

Our image series consists in all images from a given CCD, 
gathered in exposures of one of our fields in a given band. These
images are equipped with a WCS accurate to better than 1 pixel,
which can be used to associate all stars detected in the image series.
We typically have $\sim$ 200 stars, with a total of $\mathcal{O}(50000)$ measurements.
Because of saturation, the brightest stars are only usable on the poorest IQ images, while the 
faintest ones can only be measured on the best IQ images.
For the star $i$, its expected position ${\vec P_{ij}}$ in image $j$ is modeled as
\begin{equation}
{\vec P_{ij}} = T_j({\vec X_i} + {\vec \mu_i}(t_j-t_0)) \label{eq:astrom_model}
\end{equation}
where $T_j$ is the coordinate mapping from a reference system
to pixels in image $j$,  ${\vec X_i}$ refers
to the coordinates of star $i$ in this reference system, 
${\vec \mu_i}$ the proper motion of this star, $t_j$ the epoch of image $j$
and $t_0$ some reference epoch. The parameters of the astrometric fit
are $T_j$ (one per image), ${\vec X_i}$ and ${\vec \mu_i}$ (one position and one proper motion
per star). We choose the best IQ image as the reference
system: its transformation $T$ is just the identity and is not fitted.
We choose the reference epoch $t_0$ as the mean survey epoch. The transformations
$T$ are modeled as polynomial functions of the cooordinates\footnote{For sake of completeness, we note that rather than fitting $T$, we fit the reciprocal functions because the fit is then linear.}, and we chose
quadratic functions because a higher degree did not seem to improve significantly the residuals. Conversely, linear transformations increase the r.m.s
residuals by a factor of 2 to 3 with respect to quadratic ones. Note that these transformations map images from the same 
intrument on each other, and coordinate mappings from CCD coordinates to the sidereal coordinates do require higher orders.

The model is almost degenerate: if one operates the substitutions:
\begin{align*}
\hskip 2cm T_j({\vec X_i}) & \leftarrow  T_j({\vec X_i}  + (t_j-t_0) g({\vec X_i})) \\
{\vec \mu_i}    & \leftarrow  {\vec \mu_i} - g({\vec X_i}) 
\end{align*}
where $g$ is any arbitrary function, the predicted position 
is not changed if the transformations $T_j$ are linear, which
happens to be almost exactly true. 
Non-moving objects obviously lift the degeneracy,
and inserting galaxies into the fit seems an obvious cure.
However, the position of galaxies with respect to stars
depends on the chosen definition of position, and are likely
to depend on details of the PSF. We hence resorted to
iteratively isolating a subset of stars affected by proper motions:
all stars are initially fixed, and we release at most one star per iteration,
requiring that its release decreases its contribution to the $\chi^2$
by a factor of 2 or more. We also allow each iteration to discard a small
number of outlier measurements. The fit stops when no star status
was changed (fixed or allowed to move) nor any measurement was discarded.

\begin{figure}[hh]
\centering
\includegraphics[width=\linewidth]{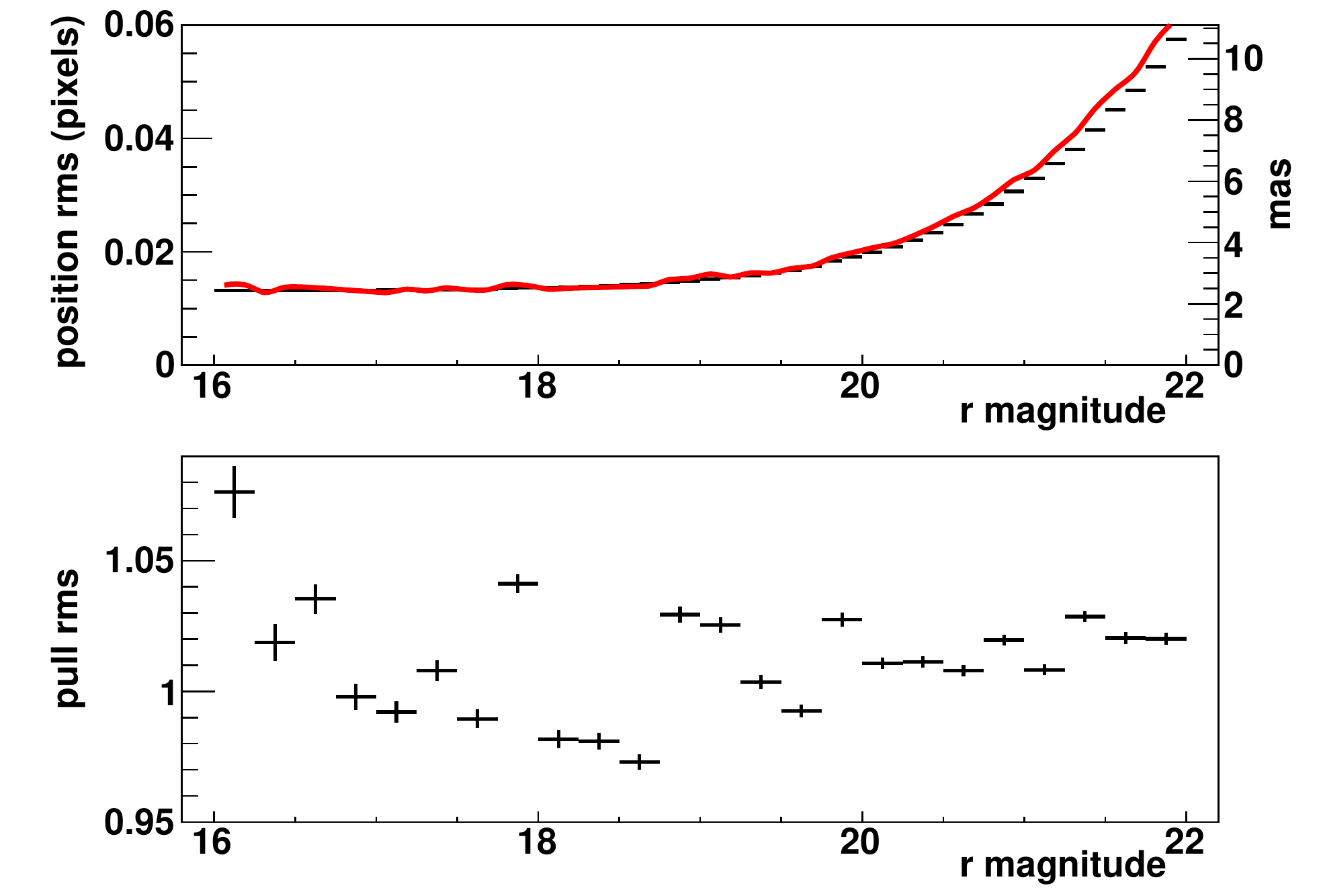}
\caption{Astrometric 1-D residuals scatter as a function of star
  magnitude for the D3 field in $r$ band. The top plot compares,
  as a function of magnitude, the
  measured residual rms (points) with the average expected rms (curve)
  including a noise floor of 0.013 pixels. They are roughly compatible, but not
  necessarily equal because the expected rms varies with IQ at fixed
  magnitude. The bottom plot displays the rms of the residual pulls
  (i.e. residuals in unit of expected rms), which are close to 1 at
  all magnitudes. We hence conclude that adding the position noise
  floor of 0.013 pixels (2.4 mas) in quadrature to the position
  uncertainty expected from shot noise fairly describes the
  residuals. This figure only considers residuals along $y$ for
  reasons explained in section \S~\ref{sec:atmospheric-refraction}.
\label{fig:astrom-resid}
}
\end{figure}

We initially used the position uncertainties from PSF position
measurements (i.e. propagation of shot noise), but those proved to be
inadequate at the bright end, where they possibly reach 0.002 pixel r.m.s.
We hence added in quadrature a ``position noise floor'' of 0.013
pixels r.m.s, adjusted to the residuals at the bright end. Figure~\ref{fig:astrom-resid} 
illustrates that this simple uncertainty model adequately describes the
observed scatter. We find similar noise floor values for all bands. 
This astrometric noise floor of 0.013 pixel or 
2.4 mas is significantly better than the one reported for
wide-field ground-based astrometry in \cite{Anderson06}. 
In contrast, \cite{Lazorenko06} obtains a significantly 
better result than ours, but on a narrower field instrument.

Although the astrometric precision has a negligible influence on the
quality of our photometric measurements, we searched for systematic
errors contributing to the astrometric precision floor, and we report
one related serendipitous finding (which does not explain the observed
residual) in appendix \ref{app:pixel-size-variations}.

The precision of proper motion measurements can be assessed by
comparing the proper motions detected in two different bands, because
the fits are performed independently. Comparing non-zero proper
motions in $r$ and $i$ bands, we find rms differences of
$\sim$1.5~mas/y per coordinate, indicating a precision of
$\sim$1~mas/y in each band. This figure is about 5 times worse than anticipated
from the single-image astrometric precision of $\sim$7 mas, and we
attribute most of this difference to our crude algorithm used to separate
fixed and moving stars, which does not aim at optimising proper motion
measurements. Optimising the proper motion precision would also
likely benefit from accounting for atmospheric refraction (that we
discuss in next paragraph) as well as pixel size discontinuities
(discussed in appendix \ref{app:pixel-size-variations}). Note that
the influence of proper motion inaccuracies on geometrical
transformations are indeed tested by the simulations described in
\S~\ref{sec:simulations}.

\subsection{Position shifts induced by atmospheric refraction}
\label{sec:atmospheric-refraction}
\begin{figure}[h]
\centering
\includegraphics[width=\linewidth]{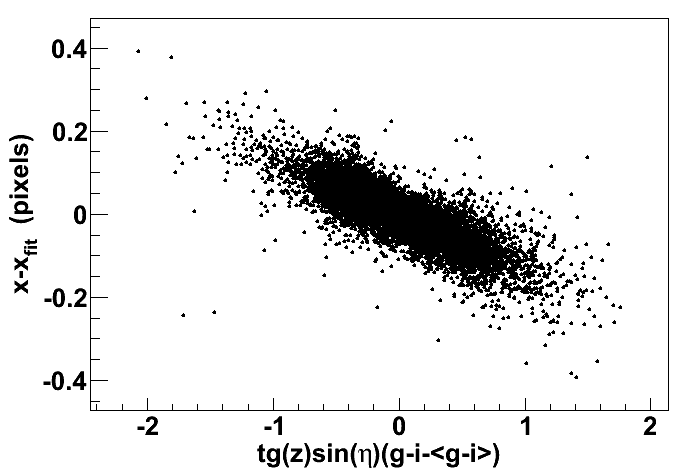}
\caption{Astrometric residuals along $x$ (i.e. RA) in $g$ band
 as a function of the displacement expected from refraction, up 
to an unknown constant, for stars brighter than $i=20$.
\label{fig:refrac_displacement}
}
\end{figure}

This section justifies why we can ignore the effect of atmospheric
refraction when mapping coordinates of different images.
Atmospheric refraction bends light rays in a plane that contains
the incoming direction and the vertical at the observatory. Light rays
from zenith are unaffected. This bending displaces the objects 
in the image plane:
\begin{align}
\delta x & = [n(\lambda)-1] \tan z \sin \eta \label{eq:deltax} \\
\delta y & = [n(\lambda)-1] \tan z \cos \eta \label{eq:deltay}
\end{align}
where $n(\lambda)$ is the refraction index of the atmosphere, 
$z$ is the zenith angle and $\eta$ is the parallactic angle, the direction
of the refraction-induced displacement in the image plane. We recall that with MegaCam, $x$ and $y$
are well aligned with right ascension and declination.
The law of sines relates the parallactic angle with other angles
describing the observing conditions:
\begin{equation}
\frac{\sin \eta}{\cos \ell} = \frac{\sin h}{ \sin z}
\end{equation}
where $\ell$ is the latitude of the observatory and $h$ is the hour
angle, i.e. the RA difference between the target and the zenith.
Atmospheric refraction displaces the whole
image (by $\sim$ 30-40\arcsec\ in the visible, for $\tan z=1$,  on the Mauna Kea), with a small distortion due to the variation of zenith angle across the field of view. Both effects are absorbed 
into the geometrical transformations (eq. \ref{eq:astrom_model}).
Conversely, we are sensitive to the different
displacement of wavelengths within the observing band, which is oriented in the same
direction as the total displacement and scales with the difference of
refractive indices between cuton and cutoff wavelengths of the band. As a consequence, 
$g$ is the most affected band. This differential displacement moves red and
blue stars in opposite directions w.r.t. an average-color star,
and we are only sensitive to the scatter of these displacements across
images, which scales with the scatter of $\tan z \sin \eta$ and $\tan z
\cos \eta$ along $x$ and $y$ respectively. It turns out that
$\sigma(\tan z \sin \eta) \sim 0.4$, and $\sigma(\tan z \cos \eta)$ is
typically 10 times smaller (see table \ref{tab:trig} below). This is a consequence of our observing
the science fields over as long a season as possible, and we will
concentrate in the following discussion on the $x$ coordinate, the most affected one. We choose to index
star colors by $g-i$, and, assuming their spectra to be power laws, we can approximate the
displacement of a star in a given image with respect to its average
position as:
\begin{equation}
\delta x_g \simeq k_g (g-i-\langle g-i\rangle) \tan z \sin \eta \label{eq:delta_x}
\end{equation}
where $k_f$ is a constant depending on the considered filter $f$, and
$g-i-\langle g-i\rangle$ is the difference in color of this star to the average
color of the stars involved in the astrometric fit. We have assumed
in equation~\ref{eq:delta_x} that $\langle\tan z \sin \eta\rangle\sim 0$, 
which is fairly accurate for a survey such as SNLS (see table \ref{tab:trig}),
and  figure~\ref{fig:refrac_displacement} illustrates that this
expression~\ref{eq:delta_x} describes a detectable effect.  We measure
$k_g \simeq 0.13$ pixels from this figure
and computations using the \cite{Pickles98} stellar library yield a similar value for
the typical Mauna Kea air column.

We evaluate the color spread of stars involved in the fit to
$\sigma(g-i) \simeq 0.9$. In the SNLS, we typically have $\sigma(\tan
z \sin \eta) \simeq 0.4$, so that refraction contributes
$\sigma(\delta x) \simeq 0.9*0.4*0.13 = 0.047$ pixels to astrometric
residuals along $x$ for $g$ band, compatible to $\sim 10 \%$ with the
difference in scatter between residuals along $x$ ($0.057$ pixels) and
$y$ ($0.026$ pixels).

We have considered incorporating differential refraction into the
astrometric model (equation \ref{eq:astrom_model}). As for proper
motions, we would then have to account for the star displacements
induced by refraction when carrying out the simultaneous photometry
fit. Our insistance on treating SNe and tertiaries as similarly as
possible would then face a problem: predicting the displacement 
requires a color, SN colors vary with phase
and we do not have color measurements at all phases. More
fundamentally, setting up a photometry scheme that relies on the
colors of the object we are trying to measure is not very
appealing. The alternative is to just ignore refraction for both stars
and SNe, and we will now evaluate the incurred loss of photometric
accuracy.

We now evaluate the effect of displacements induced by refraction on
the ratio of SN flux $f_{SN}$ to tertiary star flux $f_*$, using
expression \ref{eq:deltaF}, and averaging over tertiaries. We have:
\begin{align}
\frac{\delta E \left[ f_{SN}/f_* \right]}{ E \left[ f_{SN}/f_* \right]} & =
   \frac{\delta f_{SN}}{f_{SN}}   - E \left[ \frac{\delta f_{*}}{f_{*}} \right] \nonumber \\
   &= \frac{1}{4} \frac{  (\delta x_{SN})^2  - E [(\delta x_*)^2]}{  \sigma_{IQ}^2} \nonumber \\
   & = k_g^2 \frac{(c_{SN} - \bar{c})^2   - Var (c_*)}{ 4  \sigma_{IQ}^2} \mathrm{Var}[ \tan z \sin \eta ] \nonumber \\
    & \simeq 1.5\ 10^{-4} \left[(c_{SN} - \bar{c})^2   - 0.7 \right] 
\end{align}
where $c$ denotes $g-i$, $\bar{c}$ the average over stars, and we have
assumed $ \sigma_{IQ} = 2$. We see that in order to bias flux ratios
by more than one part in thousand, the color of the SN would need to be
3 magnitudes different from the average star. This is not the case for SNe~Ia at the
redshifts we are considering, and such sources are very rare.  Would
the algorithm have to cope with such an odd source, one could still
incorporate the displacement induced by refraction into the model.
For ``regular'' sources, and in particular SNe~Ia, it is in fact more
favourable (and obviously simpler) to just ignore atmospheric 
refraction, which will be our line of conduct in what follows.

\subsection{Photometric fit}
\label{sec:DSP-photometric-fit}
The DSP algorithm discussed in this section does not resample the images, but instead
resamples the model, so that the data pixels to which the model
is compared are indeed independent, and diagonal
least squares are not an approximation. The statistical benefit
in terms of photometric noise turns out to be small in our case
because MegaCam images are well sampled. Note that for poorly sampled
images, avoiding resampling becomes essentially mandatory for precision work.
The flux expected for image $i$ at pixel $p$ (located at ${\vec x_p}$), reads:
\begin{equation}
M_{i,p} = \left [f_i \times \phi_i({\vec x_p}-T_i[{\vec x_{obj}}]) + G\left(T_i^{-1}({\vec x_p})\right) \otimes K_i + S_i \right ] R_i \label{eq:DSP-model}
\end{equation}
where $f_i$ is the object flux in image $i$, $\phi_i$ the PSF function
in the same image, $T_i$ is the geometrical transformation to the
reference image (defined in Eq.~\ref{eq:astrom_model}), $G$ is the
galaxy pixelised model for the PSF of the reference image, and $K_i$
the convolution kernel that matches the PSF of the reference to the
one of image $i$ (at the object position), $S_i$ is the sky level in
image $i$, $R_i$ the photometric ratio of the reference to image $i$,
and ${\vec x_{obj}}$ is the coordinate of the object (in the reference
frame). $f_i$, $G$, $S_i$ and ${\vec x_{obj}}$ are the fit
parameters. The kernels $K_i$ are fitted from the PSFs of the
reference image and the image $i$, and has to be multiplied by the
photometric ratio of the same images, fitted from the PSF fluxes,
a by-product of PSF modeling (\S~\ref{sec:PSF-modeling}). Modeling
uniform photometric ratios over a CCD seems adequate because we have found
a high correlation of photometric ratios of different CCDs within an exposure.
This approximation does not cause flux biases but only contributes a small 
additional random error, bounded by the reproducibility of bright star fluxes, i.e. $\sim$0.006 mag (\S \ref{sec:photometric_uncertainty}).

The galaxy model $G$ is by default modeled at the same
  sampling as the input images, although one can consider a finer
  sampling. As written above, Megacam images are well sampled
  ($FWHM>4$ pixels on average), so that resampling does not significantly
  smooth the sharpest objects. One might be concerned because the
  galaxy is modeled at the best IQ of the image series, and the
  shortcomings of resampling under-sampled images might be at play. It
  turns out that in expression \ref{eq:DSP-model}, the galaxy model
  $G$ is smoothed by a convolution kernel. We hence resample 
  a galaxy model with the same IQ as the used images, which
  are on average well sampled. We note that the simulations we
  describe below to test the DSP method could indeed detect a bias
  induced by resampling, in particular by looking for photometric
  residuals varying with IQ, and found none.

For the fit of tertiaries, the galaxy part of the model is set to zero,
and the position in each image ${\vec x_{obj,i}} = {\vec x_{obj}} + {\vec \mu_{obj}}(t_i-t_{ref})$ 
accounts for proper motion ${\vec \mu_{obj}}$ when applicable.

\subsubsection{Fitting all bands simultaneously?}
Readers might wonder why we process pass-bands independently,
  since enforcing common values of nuisance parameters (e.g. star
  positions, proper motions) usually reduces random errors of
  parameters of interest (e.g. SN and star fluxes). Fitting the astrometry
  in all bands simultaneously would reduce the uncertainty of output
  catalogs, provided the impact of atmospheric refraction remains
  small (note that \S~\ref{sec:atmospheric-refraction} only discusses
  refraction-induced offsets between images from the same band, not
  offsets between different bands). However, the photometric fit does
  not use positions from the astrometric fit, but only uses the fitted
  proper motions and transformations. The uncertainties of the proper
  motions are too small to compromise the photometric
  accuracy. Improving the fitted catalog will marginally improve the
  fitted transformations, since their uncertainties mostly result from
  position measurement uncertainties in the image they are mapping.

Regarding the simultaneous photometric fit itself, fitting jointly all
bands would improve the quality of positions, which does not reduce
the variance of fluxes but instead reduces the bias of flux estimators
at low S/N (see \S~\ref{sec:psf-photometry}). For SNLS, this last
point is only relevant in practice for distant supernovae, which may
exhibit low S/N in $g$ and $z$ bands. Conversely, all supernovae 
have a large enough integrated S/N in $r$ and $i$ bands. We hence
fit supernovae in $g$ and $z$ bands at a fixed position, provided
by $r$ and $i$ bands, and study the possible shortcomings of the 
procedure in \S~\ref{sec:coordinate_transfer}.

Since the benefits of fitting all bands simultaneously are at best
tenuous, and might require a proper accounting of refraction, we did
not attempt it.

\subsubsection{Dealing with the Poisson noise from objects}
As discussed in \S~\ref{sec:psf-photometry}, we deliberately
ignore the contribution of the objects to the noise when
estimating fluxes, in order to ensure linearity, independently
of the fidelity of the PSF. As a consequence, the flux uncertainties
obtained from the second derivatives of the $\chi^2$ at minimum
are underestimated. The parameters, and their actual uncertainties read:

\begin{align}
  {\hat \theta} & = (A^T W A)^{-1} A^T W D \nonumber \\
  cov({\hat \theta} {\hat \theta}^T) & = (A^T W A)^{-1} A^T W cov(D D^T) W^T A (A^T W A)^{-1} \label{eq:cov_theta_full} 
\end{align}
with $\chi^2 = (A\theta -D)^T W(A\theta-D)$, and :
\begin{description}
	\item[W] is the weight matrix actually used in the fit.
	\item[D] is the data vector.
	\item[$\theta$] is the vector containing the model parameters.
	\item[A] is the design matrix, i.e. $\mathrm{E}[D] = A \theta$.
\end{description}
In standard least squares, we would have $W^{-1}=cov(D D^T)$, and
$ cov({\hat \theta} {\hat \theta}^T) = (A^T W A)^{-1}$, which is the
minimum variance bound. For reasons discussed in \S~\ref{sec:psf-photometry}, we
choose pixel weights $w_i^{-1}=Var(Sky)$. This choice leads to a
suboptimal fit, as indicated by the Gauss-Markov theorem, but we find
that the loss in precision is insignificant. Indeed, the simulations
that follow indicate an average increase of uncertainties around
$2.5\%$ above the minimum variance bound for typical tertiary stars.

\section{Validations with simulations}
\label{sec:simulations}

\subsection{Simulation goals}

The fundamental requirement of SN photometry is the preservation of
flux ratios between field stars and SN. We therefore designed a
simulation whose aim is to ensure that this ratio is maintained across
a wide range of photometric conditions. In particular we want to
ensure that :
\begin{itemize}
\item fitting a galaxy model during SN photometry does not induce any biases. Indeed, galaxy fitting is the only algorithmic difference between SN and tertiary/calibration star photometry;
\item flux ratios are properly recovered over a wide enough range of IQ and S/N;
\item after tuning some aspects of the uncertainty model, it properly describes the observed scatter.
\item sampling the galaxy model at the same spatial sampling as the images is fine enough.
\end{itemize}

\subsection{Simulation method}

The simulation consists in modifying real SNLS science images by
adding so called \textit{fake stars} to them. These fake stars are
constructed by copying and pasting image stamps of bright, high
photometric quality stars, dubbed model stars, onto a nearby galaxy
after being appropriately dimmed. We translate the model star by an
integer number of pixels before pasting, thus avoiding any
shortcomings of resampling. Note also that the time window during
which the fake SN is turned on is randomly selected. At variance with
many proposed tests of SN photometry
\citep[e.g.][]{Schmidt98,Holtzman08}, this copy-paste method is
independent of PSF modeling, astrometric mappings, and photometric
ratios between images, and hence might detect the effects of
improper estimates of these inputs.

The idea, then, is to test a photometry by its ability to reproduce
the photometric factor used to dim the model star. To construct a
light curve for each model and fake star pairing, the same procedure
is applied for each pair on a set of images. As the RSP photometry runs
on aligned images, one can translate the pixels of the model star by 
the same amount on all images and is
guaranteed to always land on the same position on the sky. For the DSP
photometry, this is clearly not the case, and we must be careful to
select unaligned (and therefore un-resampled) images that are, by
sheer happenstance, very nearly aligned up to a translation. The
underestimation of the flux as a result of a position error, for a
Gaussian PSF, is given by equation \ref{eq:deltaF}. Given a rotation
between 2 images of angle $\Delta \theta$, a relative difference in
plate scale noted $\Delta \lambda / \lambda$ and a displacement vector
$\vec{v}$ between the model and fake star, equation \ref{eq:deltaF}
can be rewritten as:

\begin{equation}
	\frac{\Delta f}{f} = \frac{1}{4} \left( \frac{||\vec{v}||}{\sigma _{seeing}} \right)^2 [(\Delta \theta)^2 + (\Delta \lambda / \lambda)^2]
	\label{eq:deltaFWCS}
\end{equation}

We use equation \ref{eq:deltaFWCS} to select bunches of consecutive
images such that they yield a difference in flux under the $10^{-3}$
level if $||\vec{v}|| = 100 ~ \mbox{pixels}$. Fake stars constructed
with a larger value for $||\vec{v}||$ are not considered in the
analysis. We indeed find un-rotated successive image bunches because
CFHT enjoys an equatorial mount and the camera (which has no rotation
capability) is usually mounted on its top end once for a whole
dark-time run. The fake stars are only pasted during these lunations,
leaving their flux at 0 for the remaining images, thereby simulating
top-hat lightcurves for these fake ``supernovae''. Note that to avoid
correlations, we only cut and paste one fake star per galaxy per
lunation. For this simulation, we use r-band images in CCD 13 of field
D1, in CCD 11 of field D2, and in CCD 12 of field D4. The chosen CCDs
are near the center of the CCD mosaic.

\subsection{Expected biases}

\subsubsection{PSF spatial variation bias}

We expect a small simulation-induced bias as a function of
displacement from model to fake star due to variations in the PSF as
we move across the image. Indeed, the fake star generation process
cuts a star with a given PSF and pastes it in a location where the PSF
is slightly different. The induced bias as a result of this is given
by:
\begin{equation}
	\frac{\hat{f}}{f} = \frac{\Sum{ij}{} PSF_{ij}(\vec{x}) PSF_{ij}(\vec{x} + \vec{v})}{\Sum{ij}{} PSF_{ij}^2(\vec{x} + \vec{v})} \label{eq:PSFbias}
\end{equation}

Because the change in the PSF model is linear by construction with
respect to position in an image, expression 
\ref{eq:PSFbias} depends linearly on $\vec{v}$. To directly observe
this bias, we run simulations with a photometric ratio of 1 and
avoid adding Poisson noise. We also compute the expected trend using
equation \ref{eq:PSFbias} for a wide range of $\vec{v}$ summed
across all images used during the simulation. The trend expected by
direct computation matches the one observed for simulations, and the
effect is clearly linear in $\vec{v}$. This bias is well below the $10^{-3}$ level
for typical $\vec{v}$ used during the
simulation. Furthermore, the bias disappears when one averages over
$\vec{v}$ directions. We hence did not take any action to
account for the PSF variation from model to fake star positions in our
simulations.

\subsubsection{Low S/N bias}
We have seen in \S~\ref{sec:psf-photometry} that PSF flux measurements 
are biased at low S/N, due to position uncertainties. When a 
common position is fitted for a source in an image series,
the bias is lower but does not disappear.

For a flux measurement on a single image $i$ of flux $f_i$, the S/N
ratio is defined simply as the ratio of $f_i$ to $\sigma(\hat{f_i})$.
For a light-curve of any shape, the least-squares estimator of 
its amplitude $A$ has a variance that satisfies:
\begin{equation}
  \frac{A^2}{\mathrm{Var}[\hat{A}]} = \Sum{i}{} \frac{f_i^2}{\mathrm{Var}[\hat{f_i}]} \label{eq:globalSN}
\end{equation}
where $f_i$ is the expected flux in each image.

In appendix B of \cite{Guy10}, it is shown that 
the bias of $\hat{A}$ follows the same law as for a single image
(described in Eq.~\ref{eq:bias_var_flux}), namely:
\begin{equation}
\frac{\mathrm{E}[\widehat{A}]}{A} \simeq \left\{ 1 - \frac{\mathrm{Var}[\widehat{A}]}{A^2} \right\} 
\label{eq:amp-bias}
\end{equation}
For the noisiest supernova observed, this is
expected to correspond to a bias of a few parts in a thousand. To make precision
tests of the photometric accuracy at low S/N we need to take into
account this bias. The photometry's ability to reconstruct the
photometric ratio will therefore be tested as a function of its S/N,
as defined in equation \ref{eq:globalSN}. To detect any remaining
bias, we fit equation \ref{eq:amp-bias} with an additional constant
offset term $b$:
\begin{equation}
  \frac{\hat{r}}{r} - 1 = - \frac{1}{(S/N)^2} + b \label{eq:SN_bias_offset}
\end{equation}
where $r$ the flux ratio used during the cut and
paste, $\hat{r}$ the reconstructed flux ratio, and 
$b$ is a free parameter.

\subsubsection{Model star correlations}

Because the same model star is reused in multiple model fake star
pairing, we take into account possible correlations induced by this
repetition and their impact on the simulation's precision. To do this,
we increase the uncertainty on the model star flux until the $\chi^2$
per degree of freedom becomes 1 when fitting equation
\ref{eq:SN_bias_offset}. We find that we must add $1 \%$ uncertainty
to the DSP fluxes of the model star, and $0.8 \%$ to the RSP fluxes.

\subsection{Simulation parameters}

\begin{figure}[h]
  \centering
  \subfloat[For real SNLS data.]{
     \includegraphics[width=\linewidth]{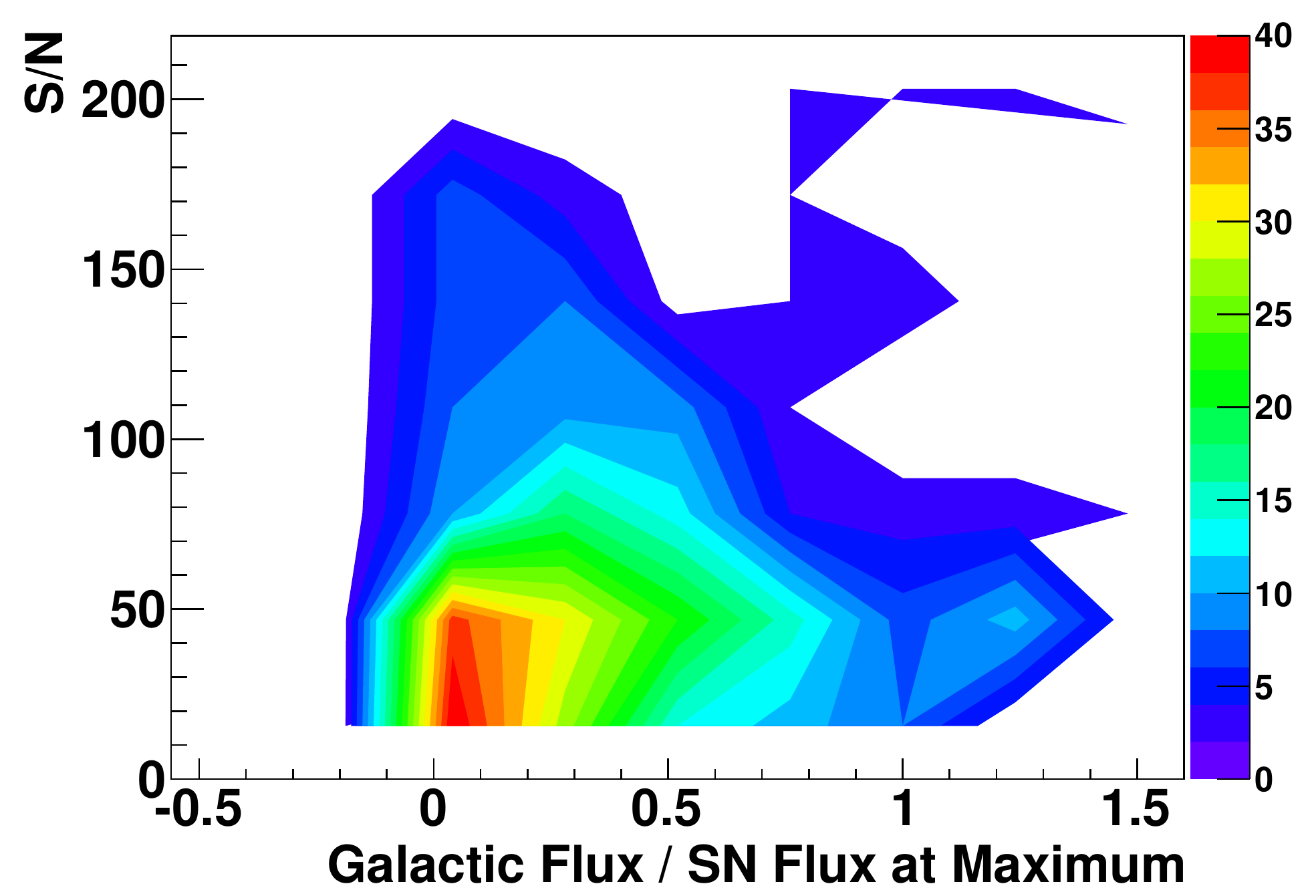}
    \label{fig:galVSsn_data}
  } \\
  \subfloat[For simulated fake stars.]{
     \includegraphics[width=\linewidth]{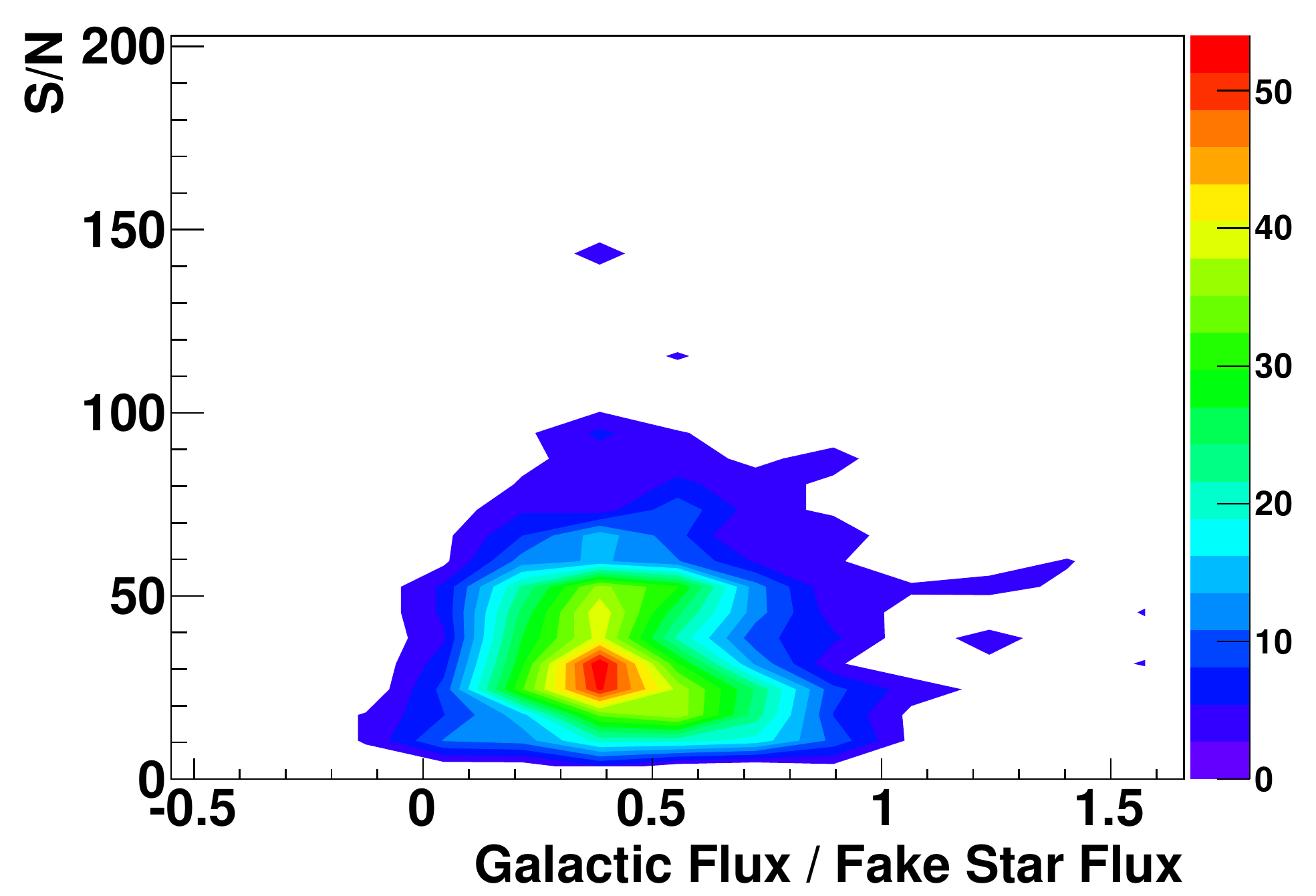}
    \label{fig:galVSsn_sim}
  }
  \caption{Above are density plots comparing the distribution of real
    SN and simulated fake stars in the plane of galactic flux VS S/N.}
  \label{fig:galVSsn}
\end{figure}

We compare the fake star's simulated parameters with those of real SN,
measured during the SNLS 3-year analysis in order to ensure that the
simulation tests the photometry in a wide range of realistic
conditions. In figure \ref{fig:galVSsn} we show density plots in the
plane of the ratio of the galaxy flux to the SN flux as a function of
the supernova S/N 
for both real data and the simulation. The galaxy flux is defined as
the integral of the galactic flux weighted by the PSF. For a
galaxy model $G(i,j)$, this is computed as :

\begin{equation}
  F_{gal} = \frac{\sum _{i,j} PSF_{\vec{x}_{SN}}(i,j) \times G(i,j)}{\sum _{i,j} PSF_{\vec{x}_{SN}}(i,j) \times PSF_{\vec{x}_{SN}}(i,j)}
  \label{eq:galflux}
\end{equation}

With this comparison, we see that the distribution of simulated
parameters resembles that of real data, however with more galaxy
flux on average in simulations than in real data. This helps at
detecting possible shortcomings of fitting a structured galaxy. 
We recall that the images used
during fake star photometry are the same as those used for SNLS
science photometry, and we can therefore be confident that the simulation
closely mimics the observing conditions of SN photometry.

In addition to selection factors that are aimed at mimicking the SN
population, we perform cuts necessary for proper analysis of the
simulation results. A number of model stars used turned out to be
variable stars. These are cut from the analysis. We also cut all fake
stars generated using a photometric factor above 0.1 so that the
original Poisson noise of the model star becomes negligible compared
to that added to the fake star during the cut and paste. Finally,
model stars that are cataloged as having a significant proper motion
are also cut, because the DSP photometry will take into account their
motion but not that of the corresponding fake SN.

\subsection{Results}

\subsubsection{Photometric accuracy of RSP}
\label{sec:RSP_accuracy}
\begin{figure}[h]
  \centering
  \subfloat[For a wide range of S/N ratios]{
     \includegraphics[width=\linewidth]{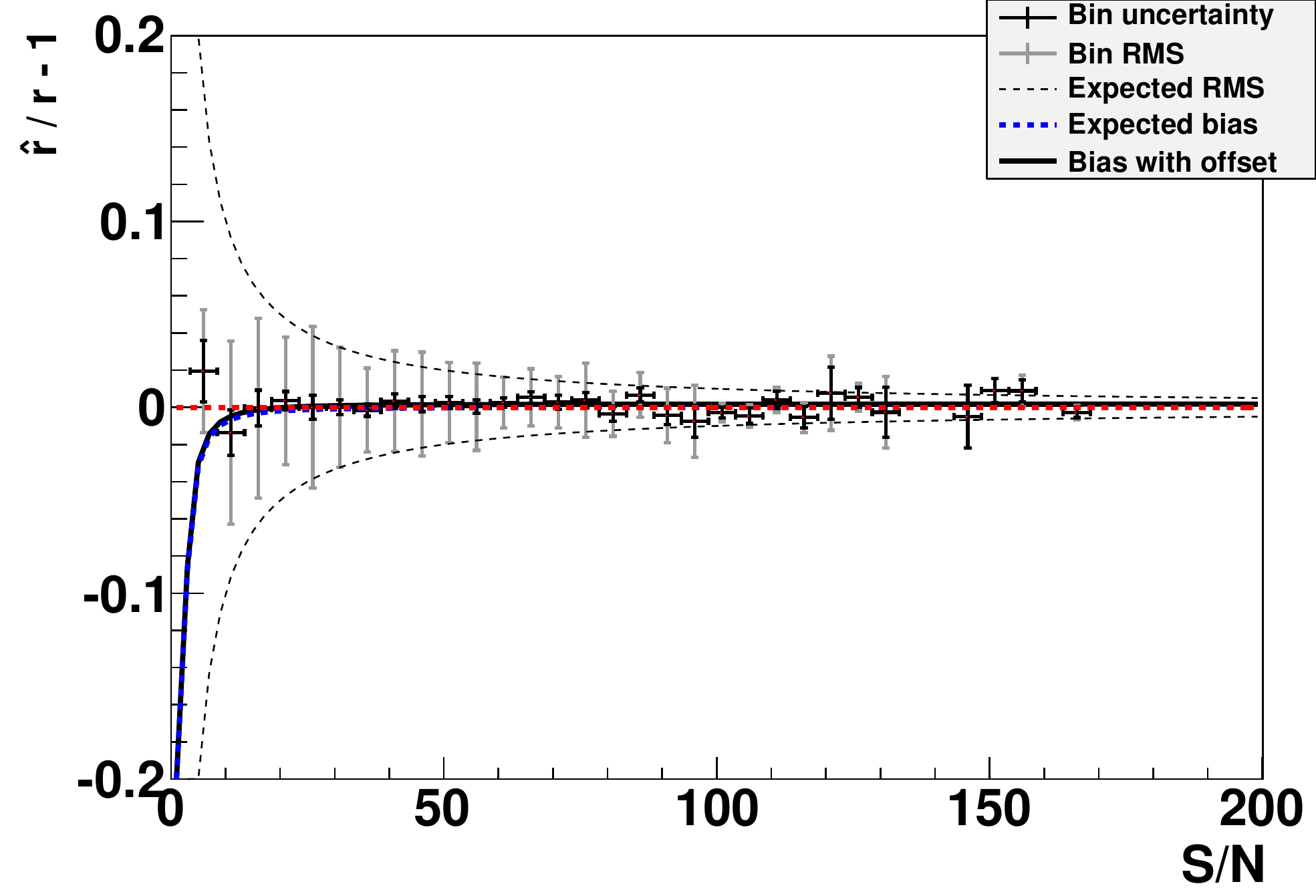}
    \label{fig:a06snplot-global}
  } \\
  \subfloat[Zoom on low S/N region]{
     \includegraphics[width=\linewidth]{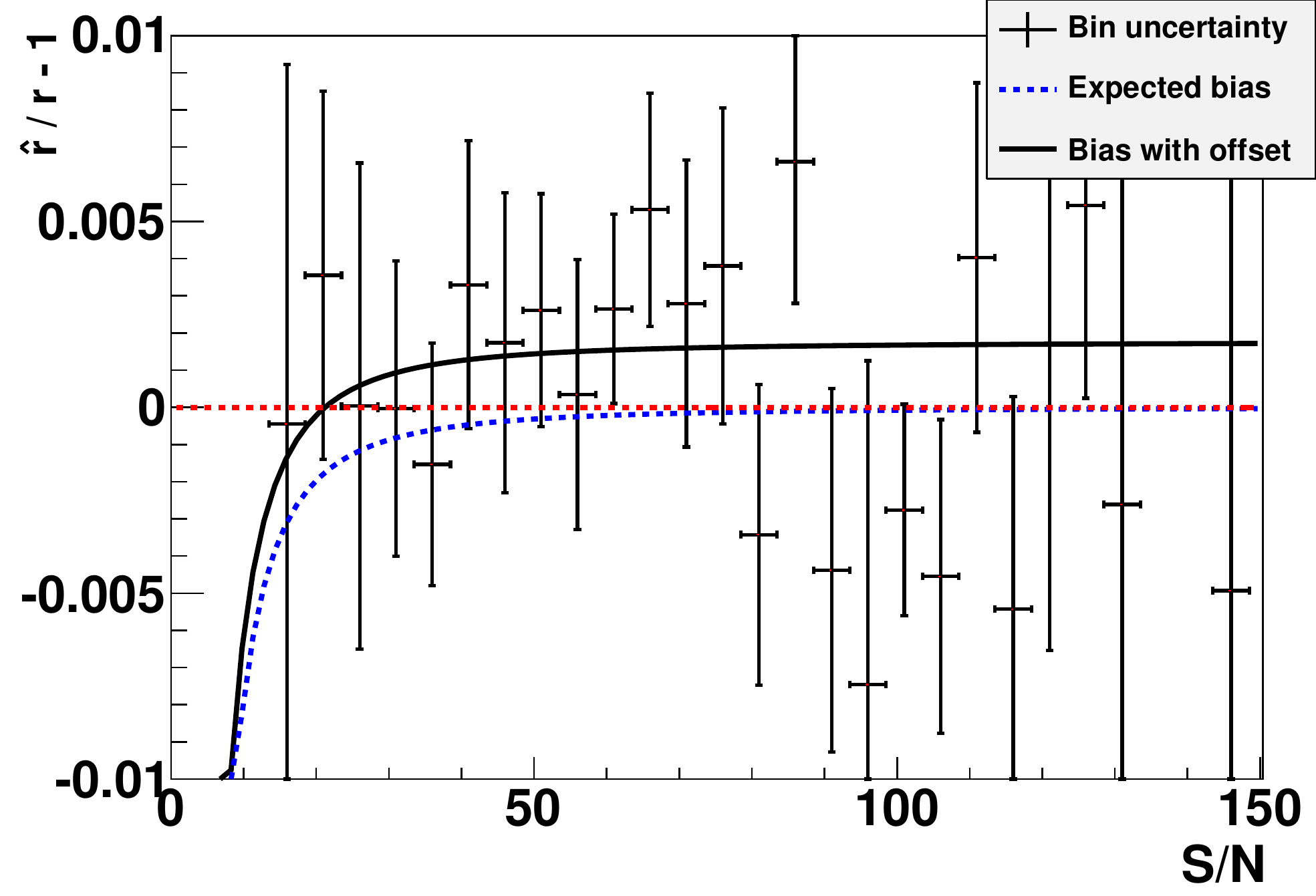}
    \label{fig:a06snplot-zoom}
  }
  \caption{Photometric factor accuracy as a function of S/N for the RSP method.}
  \label{fig:a06snplot}
\end{figure}

We begin by analyzing the results for the RSP. This technique
was used for measuring the SNLS supernovae reported in \cite{Astier06} and
\cite{Guy10}. In figure
\ref{fig:a06snplot}, we see the result of fitting equation
\ref{eq:SN_bias_offset} to the photometric ratios obtained. We find
that this method overestimates the flux of SN by a factor of $(1.75 \pm
0.83) \times 10^{-3}$. This bias has not been found to depend on galactic
flux, model or fake star flux, star color, or IQ. A number of tests
were performed in an attempt to determine its origin. These are:
\begin{itemize}
\item Reducing the vignette size used by RSP. A change would indicate
that pollutions in the vignette are causing the flux bias, but the bias remained.
\item Keeping the photometric factor at 1, and pasting the fake star
  on a dark patch of sky. We then compare the photometry of the fake
  star with and without a galaxy fit. A difference would indicate that
  flux transfers between fitted galaxy and fitted SN are causing the bias.
  No significant difference was observed.
\item Fitting the flux average of the RSP fake star lightcurve using
  the covariance matrix produced by DSP, to see if the error model of
  RSP was biasing. The bias remained.
\item Switching to $i$ band. Again, the bias remained.
\end{itemize}

We conclude that the measured bias is likely to be a statistical
fluctuation of the simulations at the $2$ $\sigma$ level. This
photometry method was used in particular to produce the SNLS light
curves published in \cite{Guy10}, and we recommend adding a correlated
$1.75\times 10^{-3}$ relative systematic uncertainty to this data set,
which amounts to less than 1/3 of the photometric calibration
uncertainty.

\subsubsection{Photometric accuracy of DSP}
\label{sec:DSP_accuracy}
\begin{figure}[h]
  \centering
  \subfloat[For a wide range of S/N ratios]{
     \includegraphics[width=\linewidth]{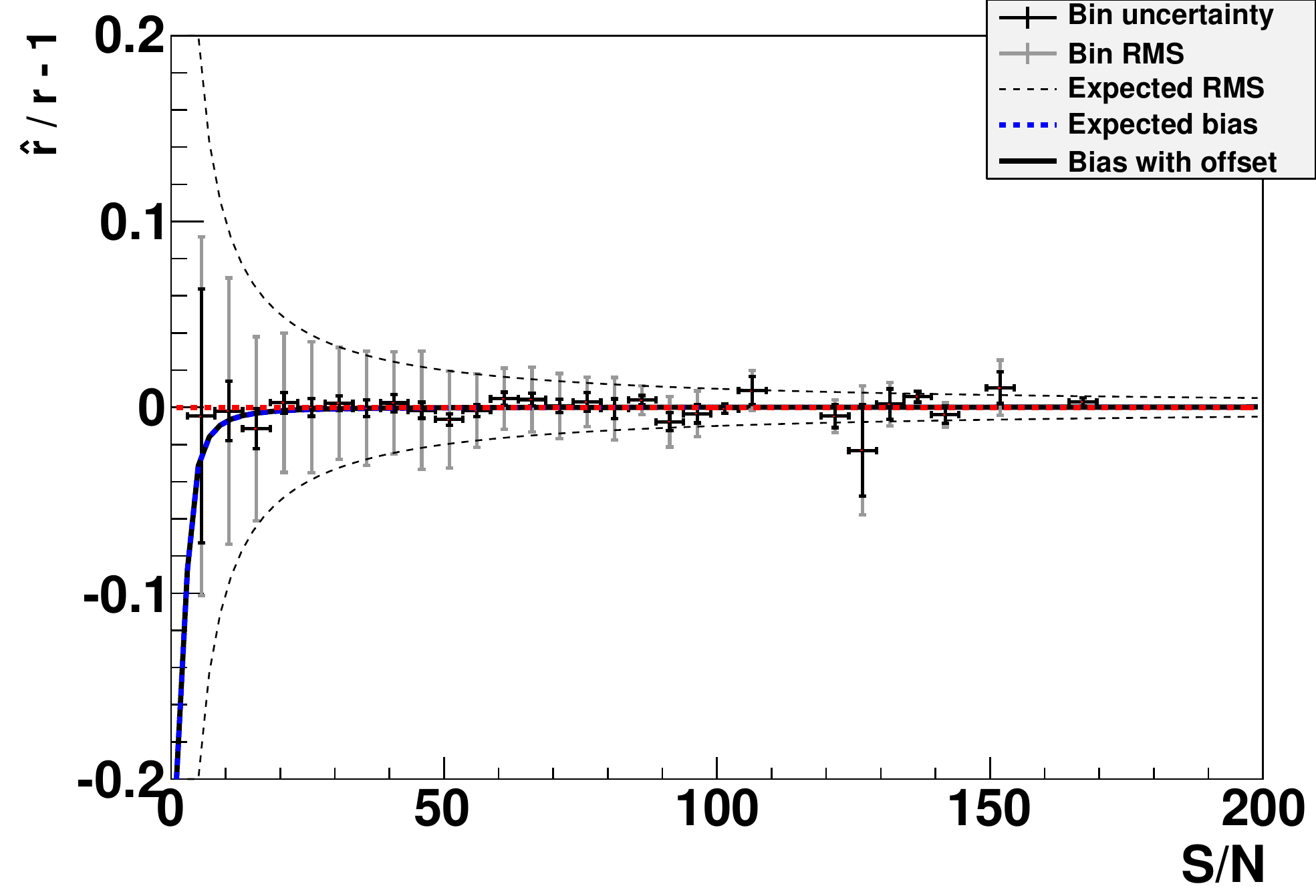}
    \label{fig:wnrsnplot-global}
  } \\
  \subfloat[Zoom on low S/N region]{
     \includegraphics[width=\linewidth]{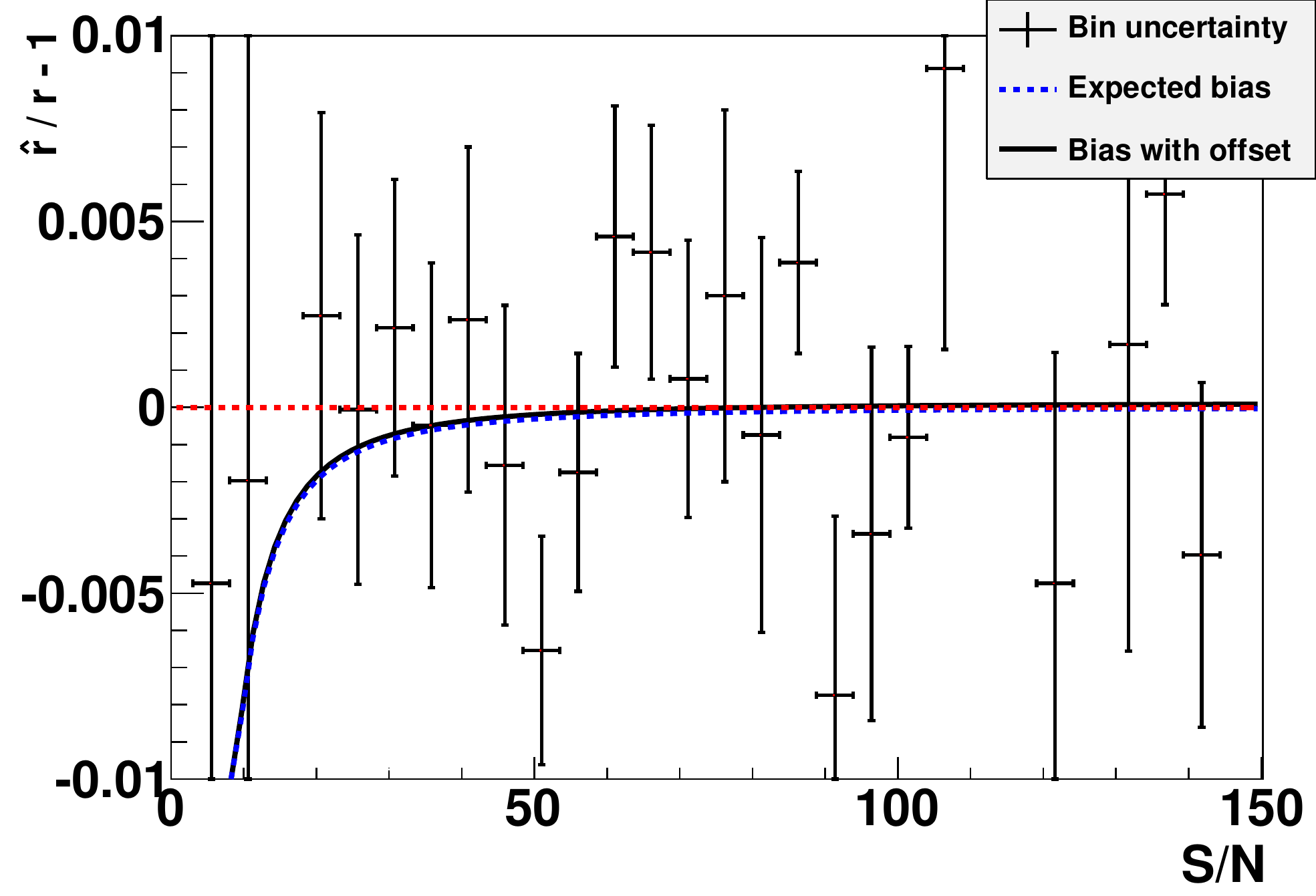}
    \label{fig:wnrsnplot-zoom}
  }
  \caption{Photometric factor accuracy as a function of S/N for the DSP method.}
  \label{fig:wnrsnplot}
\end{figure}

From this section on, the results refer to those obtained using the
DSP method. We begin by fitting equation \ref{eq:SN_bias_offset} to
the data. The fit is seen in figure \ref{fig:wnrsnplot}. We find that
no offset exists beyond the $10^{-3}$ level. The fitted offset value is
$(0.12 \pm 0.9) \times 10^{-3}$.

\begin{figure}[h]
  \centering
   \includegraphics[width=\linewidth]{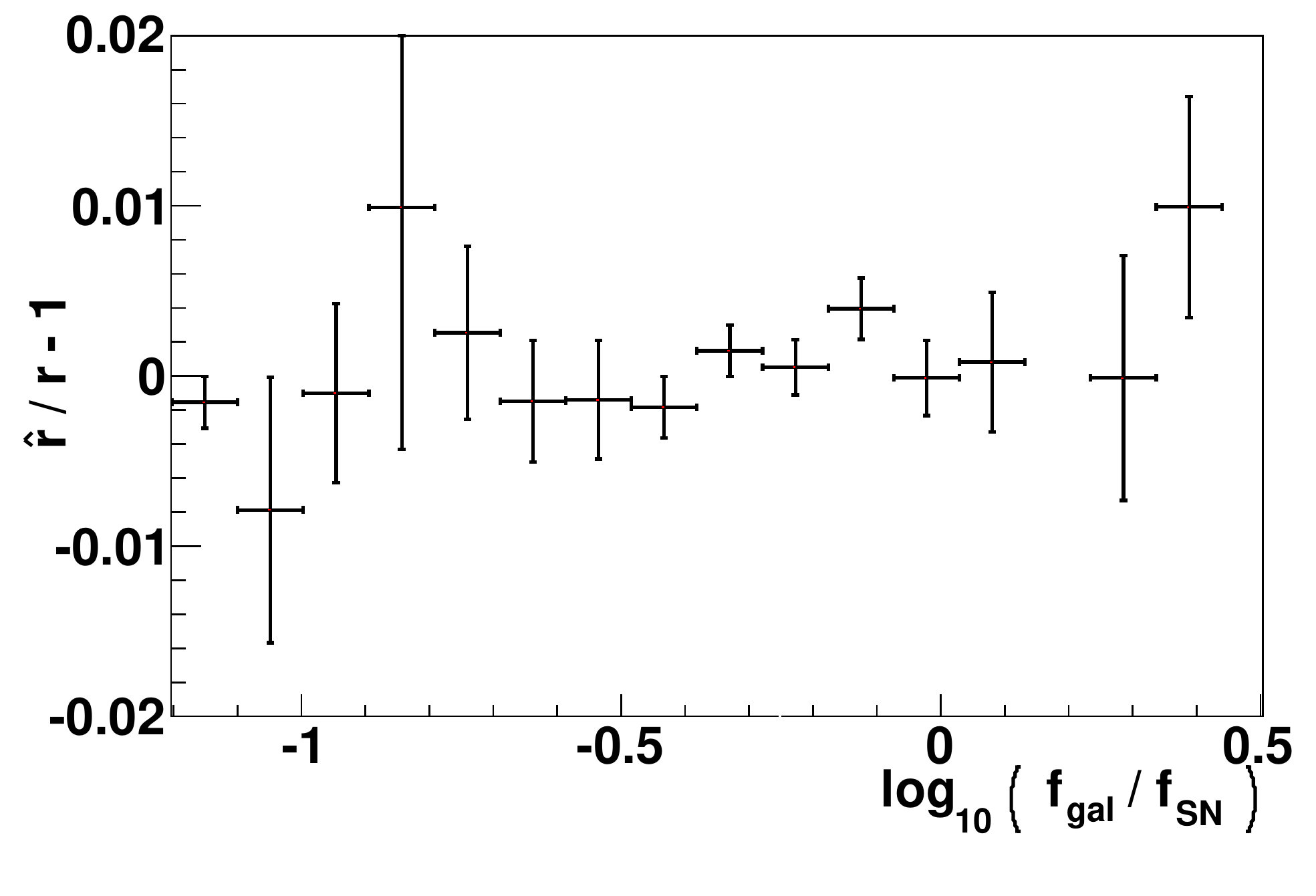}
   \caption{Here we consider photometric accuracy as a function of
     galactic flux. No significant bias is observed.}
  \label{fig:galevo}
\end{figure}

Field star and SN photometry differ most crucially in that during the
SN fit we also fit a galaxy model. We therefore also investigate
photometric accuracy as a function of galactic flux, as defined in
equation \ref{eq:galflux}. In figure \ref{fig:galevo} we look at the
evolution of photometric accuracy as a function of galactic flux
\textit{after} we have corrected for the S/N ratio bias. No
significant remaining bias is observed.

\begin{figure}[h]
  \centering
   \includegraphics[width=\linewidth]{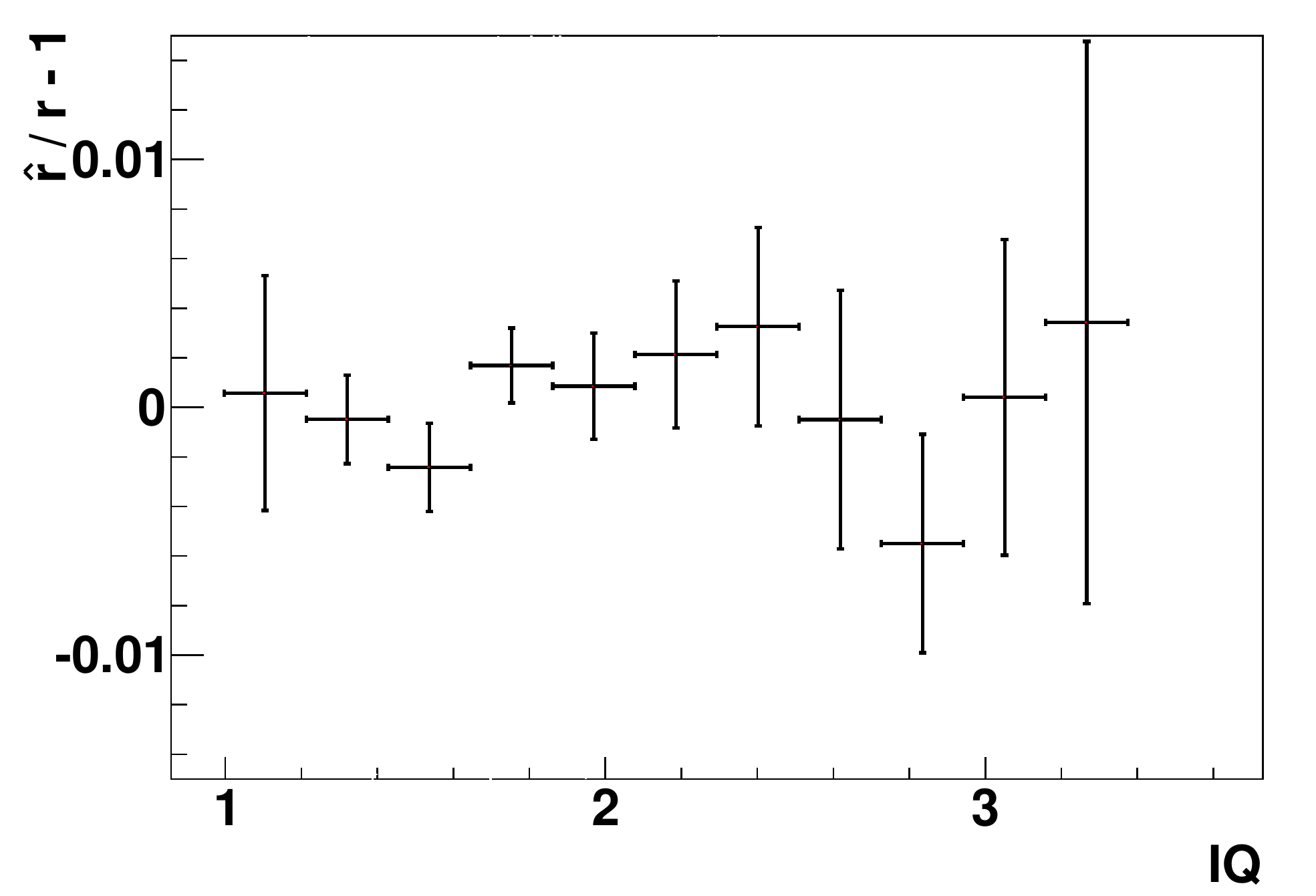}
   \caption{Here we consider photometric accuracy as a function of
     image quality. No significant bias is observed.}
  \label{fig:seeingevo}
\end{figure}

Finally, we also find that preservation of flux ratios does not vary
with image quality, as shown in figure \ref{fig:seeingevo}. Again, the S/N
ratio bias is corrected prior to investigating any bias as a function
of IQ.

\subsubsection{Photometric uncertainty results}
\label{sec:photometric_uncertainty}
\begin{figure}[h]
  \centering
  \includegraphics[width=\linewidth]{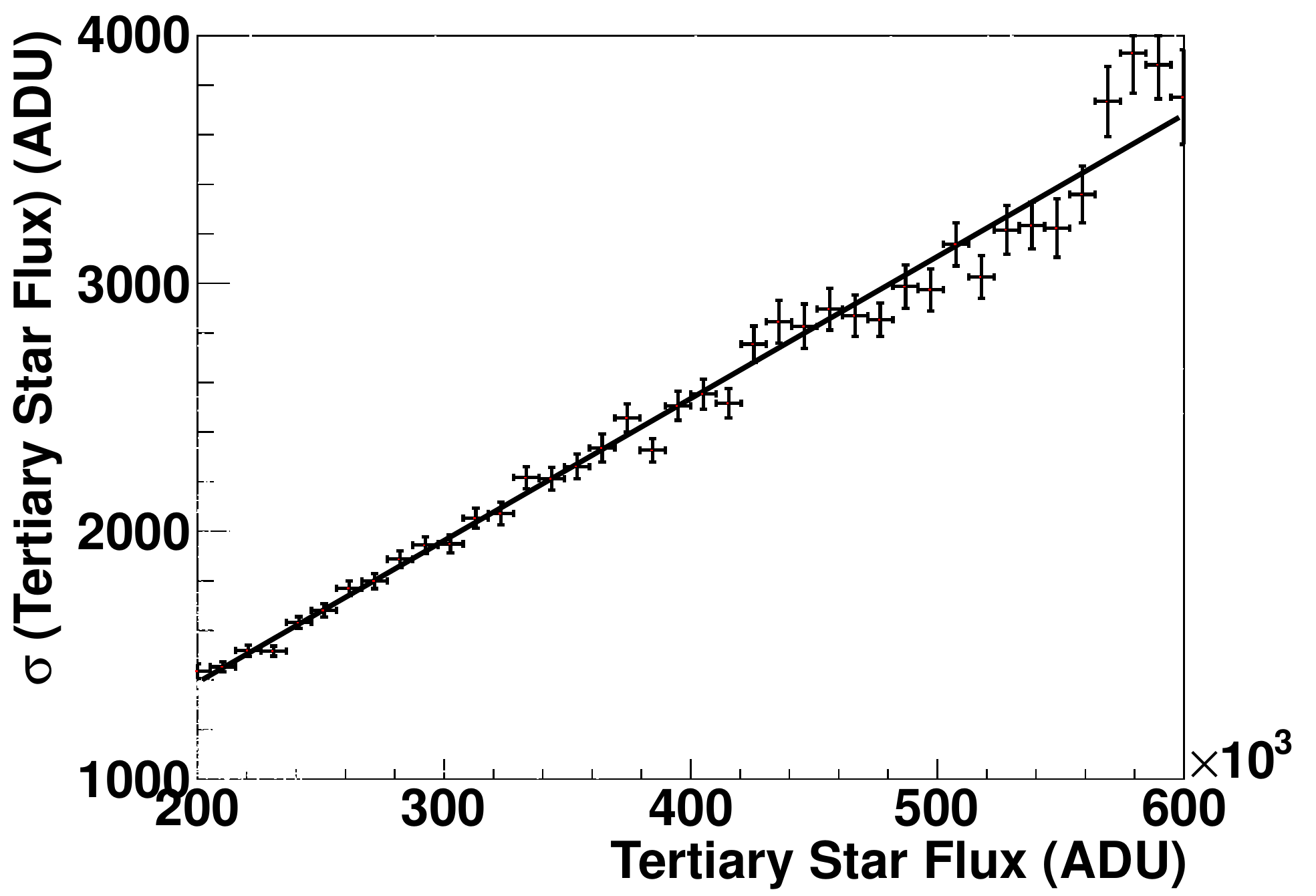}
  \caption{Standard deviation of flux estimates over the lightcurve
    as a function of the average flux. The relation is shown
    here for high-flux field stars, and we see a clear linear
    relationship, indicative of a contribution to scatter beyond shot noise
    from the sky and the object.}
  \label{fig:betafit}
\end{figure}

The output covariance matrix includes the Poisson noise of the sky and
the signal itself (both star and galaxy if present). We assume that
there exists an additional quadratic term which describes
contributions to the variance coming from errors in the PSF model, the
photometric ratio, and/or the residual photometric non-uniformity in
the images. The variance therefore takes the form:
\begin{equation}
V_{flux} = V_{sky} + \frac{1}{G} F + \beta ^2 F^2 \label{eq:errormodel} \\
\end{equation}
where G is the gain in $e^-$ per ADU.
To estimate the value of $\beta$, we use the photometry of bright
(non variable) tertiary stars. For such stars, we assume that the
$\beta ^2 F^2$ term dominates the variance. Fitting a linear relationship
between the RMS of a high flux light curve and its average flux should
therefore yield the value of $\beta$. In figure \ref{fig:betafit}, we
see the result of the fit, for which we obtain $(5.6 \pm 0.1) \times
10^{-3}$. This is essentially identical to the repeatability of 6 mmag
for aperture measurements on the same data set reported in \S~4.1 of 
\cite{Betoule13}. We are then tempted to attribute most of this 
noise floor to flat fielding rather than photometry techniques.

\begin{figure}[h]
  \centering
  \includegraphics[width=\linewidth]{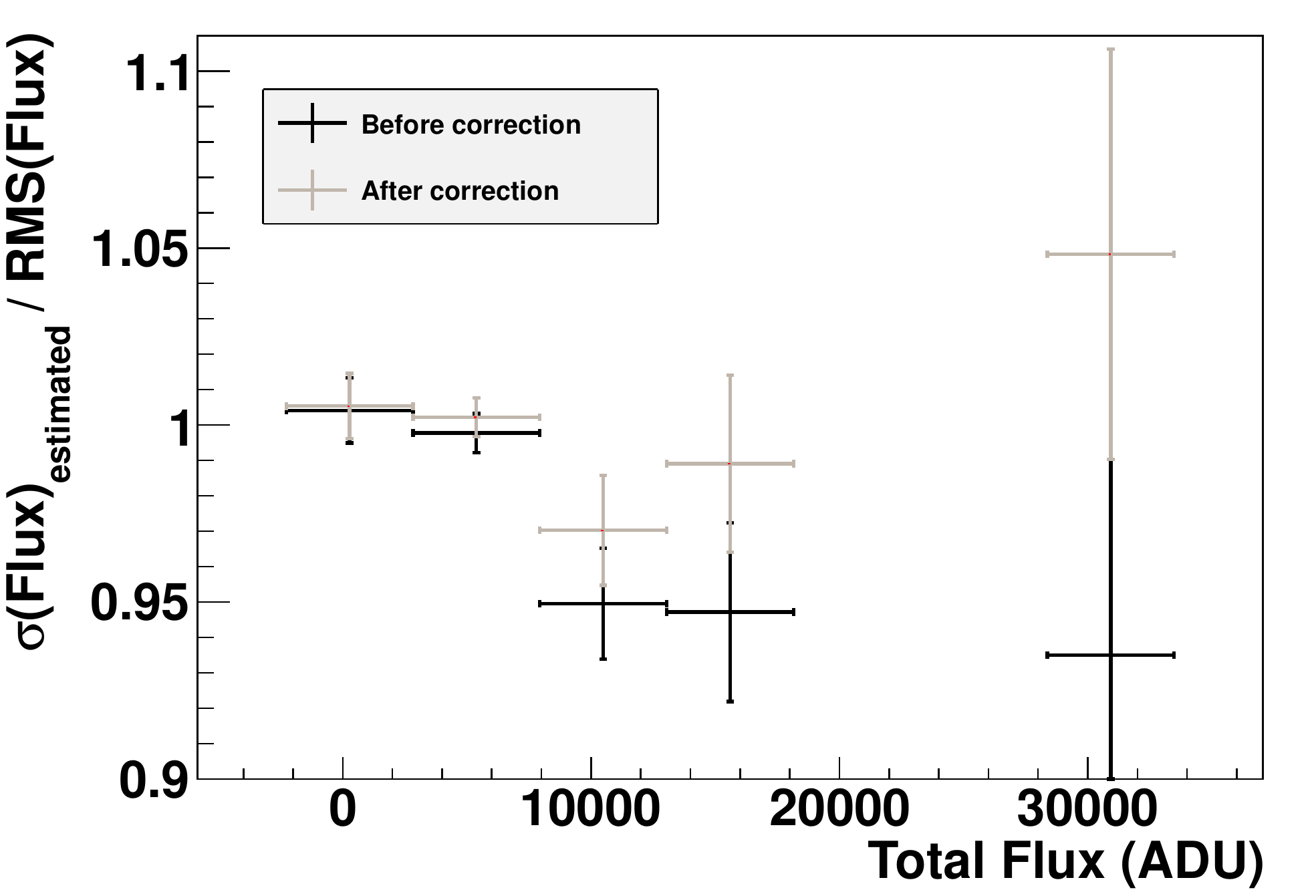}
  \caption{We plot here the ratio of the modeled
    uncertainty to the RMS of the light curve, as a function of the sum of the fluxes of the fake star
    and galaxy. The two set of points refer to before and after adding
    a $\beta$ term to the model uncertainty (Eq. \ref{eq:errormodel}). We see that the
    correction makes only a small difference.}
  \label{fig:betafake}
\end{figure}

We check that the fitted value is accurate in the low flux regime of
the fake SN. To do so, we compare the squared RMS of their light
curves with the estimated variance before and after adding a quadratic
correction. Note that because the galaxy flux contributes to the
variance, we look at the evolution of the variance as a function of
the sum of the fake star and galaxy fluxes. This is seen in figure
\ref{fig:betafake}. We see that the fitted quadratic term is
compatible with data at these low fluxes, but is also almost
negligible for such dim objects.

\subsubsection{Position reconstruction results}

\begin{figure}[htb]
  \centering
  \includegraphics[width=\linewidth]{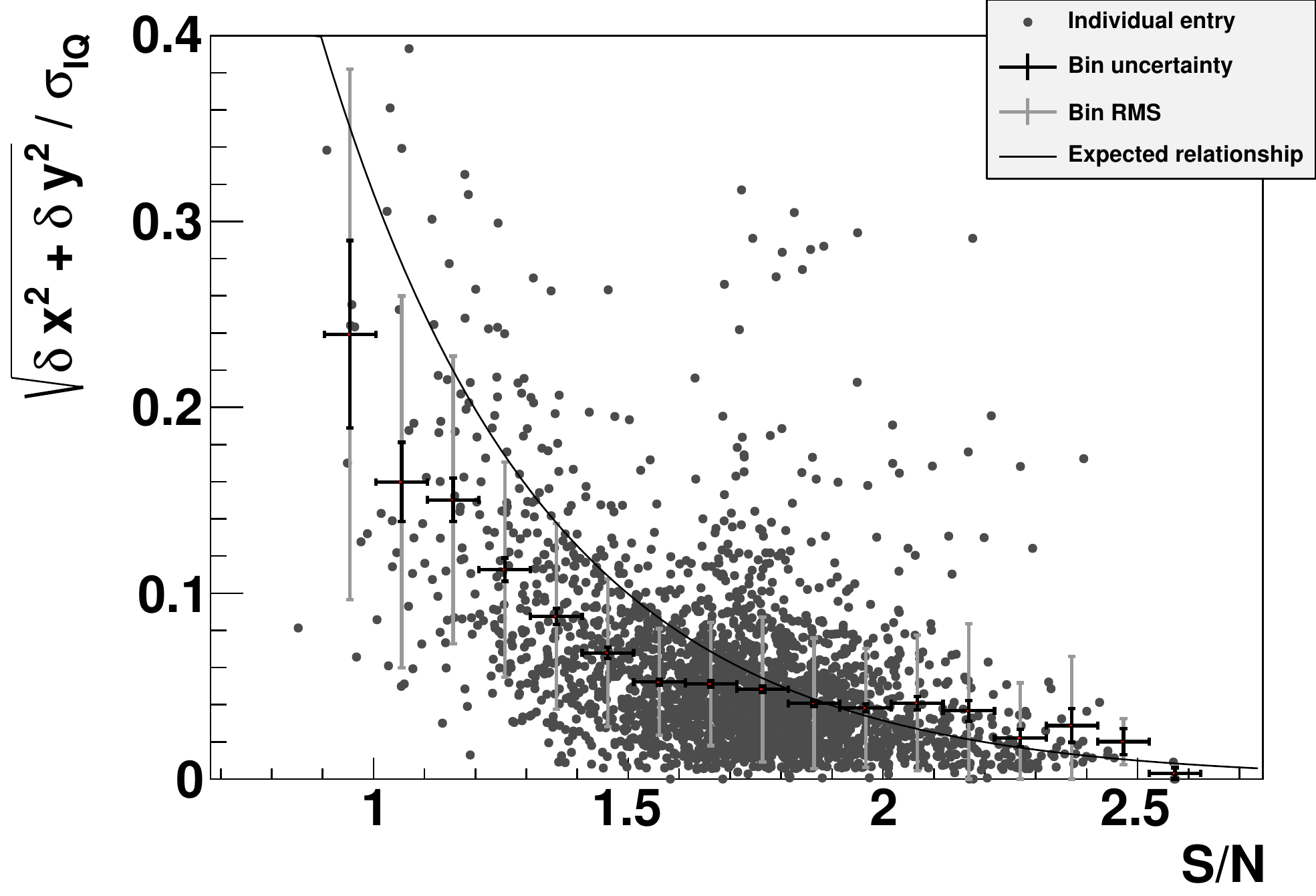}
  \caption{We consider the error in the fitted position in units of $\sigma_{IQ}$
    as a function of the S/N ratio. The discrepancy between the
    expected relationship (solid line) and observed values (points with error bars) is due to the fact that
    the form factor model does not accurately capture the non
    gaussianities of the PSF. This is explained in detail in appendix
    \ref{sec:formfactor}. The outliers observed are thought to be due
    to model stars affected by proper motions not
    detected by simultaneous astrometry.}
  \label{fig:poserrevo}
\end{figure}

By assuming that the position fit of the model star is perfectly
accurate, we can conclude that the actual position of the fake star is
that of the model plus the displacement vector used during fake star
construction. We are therefore able to compare the fake star's fitted
position with what we can reasonably assume is the correct one. In
figure \ref{fig:poserrevo}, we plot the ratio of the error on position
computed in this way to the average seeing in the image sample in
which the fake star was generated versus the S/N ratio. We also plot
the expected relationship between the two as explained in appendix
\ref{sec:formfactor}.
We notice that while the position errors follow the expected trend, a
few outliers remain. We attribute these to be due to model
stars affected by significant proper motions that have not been flagged as moving 
during simultaneous astrometry.

\section{Calibration using field stars}
\label{sec:calibration-using-field-stars}
\subsection{Calibration scheme}

The magnitudes delivered by the \cite{Betoule13} catalog, referred to
as $m_{APER}$, are the result of a comparison of tertiary star
aperture fluxes with those of standard stars. Because SN photometry is
done using PSF photometry, computing a zero point requires comparing
PSF and aperture fluxes for all tertiary stars. The PSF photometry
zero point for each image stack (all images in the same field, band,
and CCD) is given by :
\begin{equation}
  zp = \left\langle m_{APER} + 2.5 \times \log_{10} \left( \hat{f}_{PSF} \right) \right\rangle_{\mbox{all field stars}}
  \label{eq:zp_def}
\end{equation}
where $\hat{f}_{PSF}$ is the average flux over all images in the
stack. The linearity of the PSF fluxes has been shown using the
simulations discussed previously. However, the calibration of the
science images relies on the comparison of PSF and aperture fluxes. To
ensure the linearity of the entire calibration process, we must ensure
the linearity of the ratio between these two. In this section, we
discuss two significant biases in the ratio of $\hat{f}_{PSF}$ to
$\hat{f}_{APER}$, and what has been done to resolve them :

\begin{itemize}
\item First, aperture photometry makes no attempt at accounting for
  sky subtraction residuals. In other words, the obtained flux will be
  artificially modified by the total flux contribution of sky
  residuals in the aperture. On the other hand, PSF photometry actually
  fits the remaining sky level integrated by aperture photometry (this is the $s_i$ term of equation
  \ref{eq:DSP-model}). Imperfections of sky subtraction lead to a bias, particularly
  significant at low fluxes.
\item Secondly, PSF photometry does not take into account the
  chromatic dependence of the PSF. The model assumes that the PSF is
  the same for all stars, regardless of color, which obviously leads
  to a color dependency in the produced PSF fluxes. Because aperture
  photometry does not depend much on the shape of the PSF, it is not
  affected by this. This also leads to a significant difference.
\end{itemize}

\subsection{The effects of sky subtraction residuals}
\label{sec:sky_subtraction}

The sky level subtracted from the images is obtained using the average
of the image pixels computed over all pixels, except for masks placed
over all detected objects (see \S\ref{sec:image-catalog}). Despite
these masks, residual contamination from the tails of the flux
distribution of bright objects affect the remainder of the image. We
note that a prominent criteria in the selection of tertiary
calibration stars is their level of isolation, and they will therefore
be systematically less contaminated than the average pixels over which
the sky level was computed. This is why residual sky level \textit{at
  the position of tertiary calibration stars} does not average out to
0 when averaged over all images for any given star. Note that such an
effect manifests itself as a flux dependent bias because the same
residual sky level will affect the ratio of $\hat{f}_{PSF}$ to
$\hat{f}_{APER}$ more significantly for lower fluxes. This is seen
clearly in figure \ref{fig:magbias_i}. It is possible to compute the
expected residual effect by comparing the PSF tail pollutions expected
at the average distance from the nearest bright objects for tertiary
stars to the average PSF tails pollutions over the pixels used to
compute the sky level \citep[see fig. 17 of][]{Betoule13}. Indeed, the calibration catalogs given employ
such a correction.

Such a correction provides only a crudely averaged estimate of the
effect. Indeed, this only produces one single correction to be applied
to all stars equally. During the calibration process described here,
we undo this correction \citep[using the figures provided in \S 4.3.4 of][]{Betoule13} and implement our own. By using the
fitted sky level of PSF photometry, we can instead provide one
correction per star. However, such a correction only makes sense if we
can reasonably believe that the fitted sky level actually corresponds
to the left-over sky level. We expect for the fit to make up for
errors in the PSF model by artificially altering the fitted sky level
with a fraction of star's actual flux while the actual sky level 
does not scale with the star flux. To allow for a chromatic
component to PSF modeling errors (discussed in \S\ref{sec:psf_chromaticity} and \S\ref{sec:brighter_fatter}), we model the fitted sky level as:

\begin{eqnarray}
  \hat{s} &=& [a + b \times (g - i) ] \hat{f}_{PSF} + \hat{s}^{\prime} \label{eq:sky_model} \\
  \hat{s} &\mbox{:}& \mbox{is the average fitted sky level} \nonumber \\
  g \& i &\mbox{:}& \mbox{are the $g$ and $i$ magnitudes respectively} \nonumber \\
  \hat{s}^{\prime} &\mbox{:}& \mbox{is the new estimate of the average sky level} \nonumber \\
  a ~ \& ~ b &\mbox{:}& \mbox{are fitted parameters} \nonumber
\end{eqnarray}

The $a$ term is meant to accommodate both that colors are
arbitrarily defined, and that achromatic PSF errors cause transfers
between object flux and sky level. By definition, $\hat{s}^{\prime}$
is expected to be the true sky level and hence not to scale
with $ \hat{f}_{PSF}$. So,  $\hat{s}^{\prime}$ becomes negligible 
at the high-flux end of our stars, and we fit  $a ~ \& ~ b$
in this regime, thus rewriting definition \ref{eq:sky_model} as:

\begin{equation}
  \frac{\hat{s}}{\hat{f}_{PSF}} = a + b \times (g - i)
  \label{eq:sky_approx}
\end{equation}

\begin{figure}[h]
  \centering
  \includegraphics[width=\linewidth]{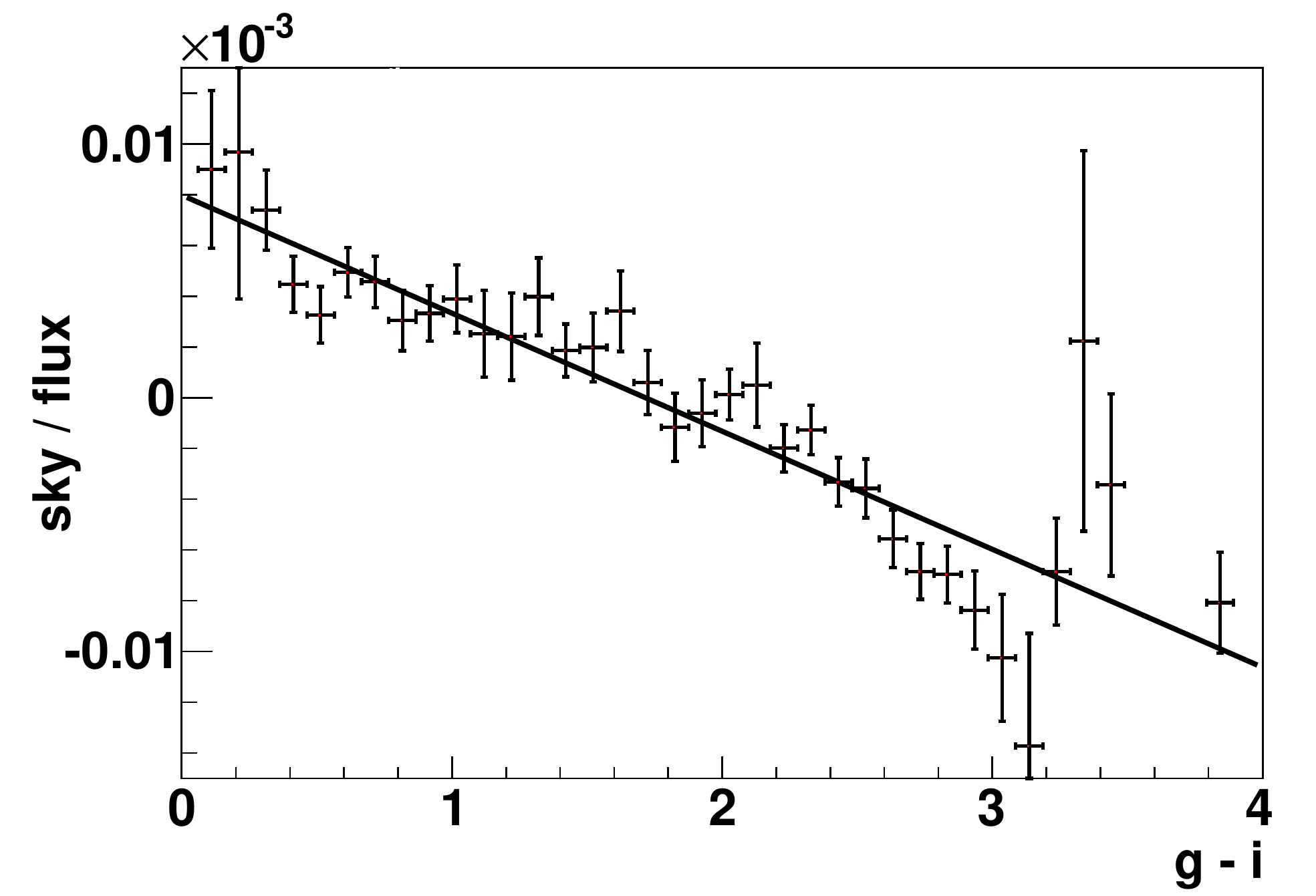}
  \caption{Ratio of the fitted sky level to the
    flux of the star as a function of color, for high flux stars only in
    $i$ band. For such stars we assume that the fitted sky level is
    predominantly a fraction of the flux incorrectly fitted as the sky
    level. We see that the fraction of flux that goes into our sky
    level estimator evolves linearly with color.}
  \label{fig:skyfluxfit}
\end{figure}

\begin{figure}[h]
  \centering
  \subfloat[Before correction.]{
    \label{fig:magbias_i}
    \includegraphics[width=\linewidth]{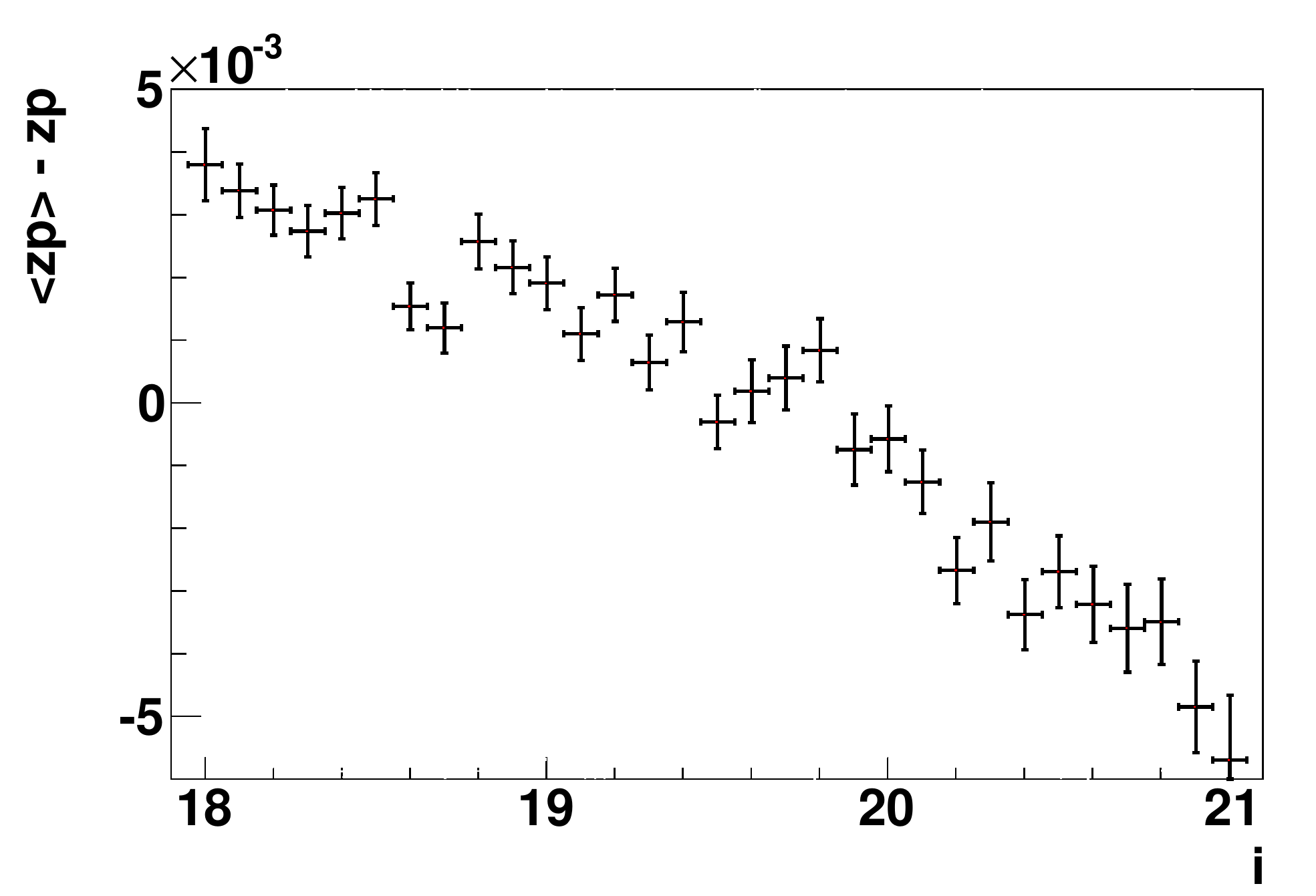}
  } \\
  \subfloat[After correction.]{
    \label{fig:magflat_i}  
    \includegraphics[width=\linewidth]{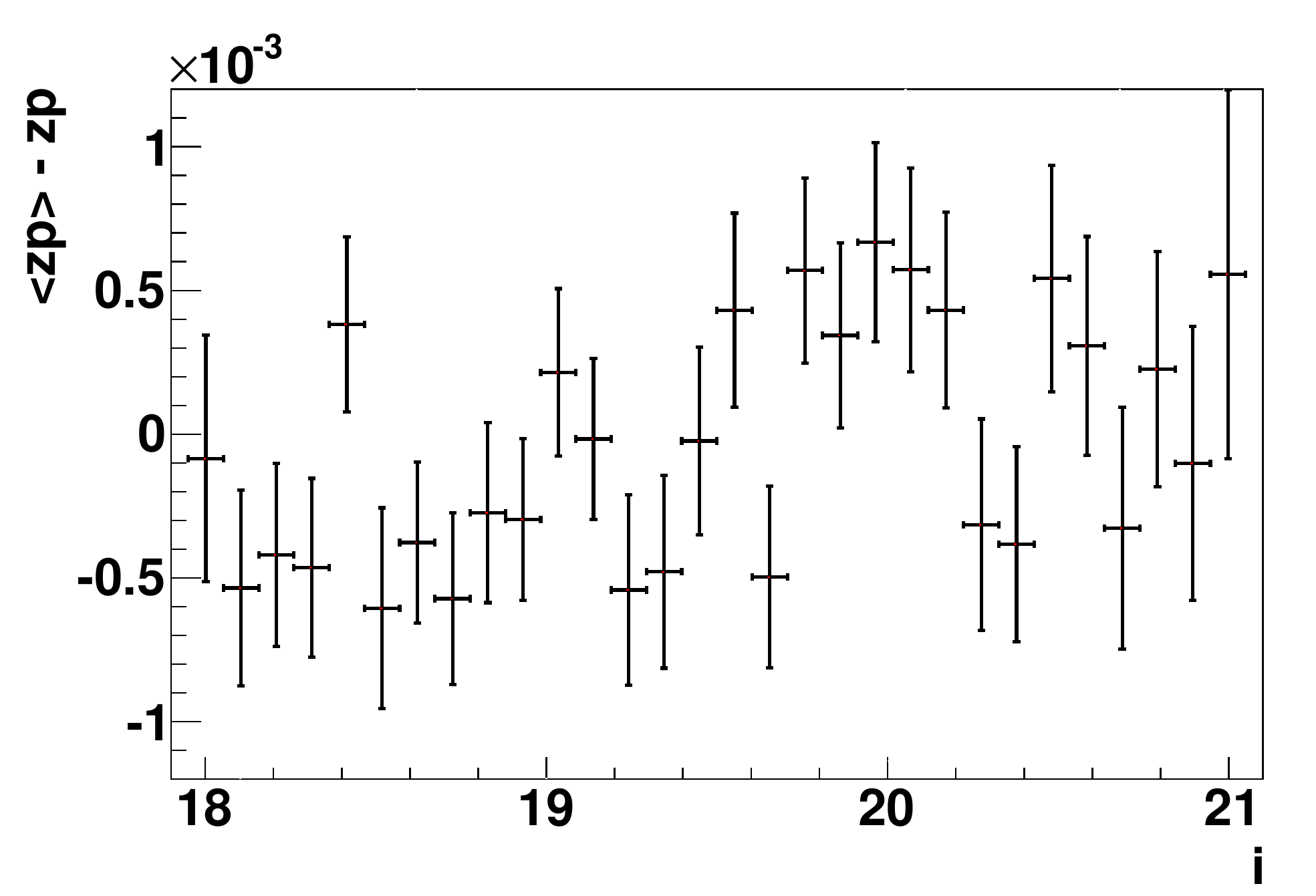}
  }
  \caption{Plot of zero point residual vs magnitude in $i$ band, before and after correcting for aperture sky pollutions.}
  \label{fig:zp_VS_mag_i}
\end{figure}

The fitted relation is illustrated in figure \ref{fig:skyfluxfit}
and the $a ~ \& ~ b$ values for each band are displayed in table
\ref{tab:skycor}. 

\begin{table}[htb]
\centering
\caption{Parameters relating fitted sky level and flux as a function of color
for bright stars (Eq.~\ref{eq:sky_approx}).
}
\begin{tabular}{ r | r @{$\pm$} r r @{$\pm$}r |}
Band & \multicolumn{2}{c}{$a \times 10^{6}$} &  \multicolumn{2}{c|}{$b\times 10^{6}$} \\ \hline
$g$ &  14   & 0.7  &  -13   & 0.5  \\
$r$ &  8.4  & 0.3  &  -5.4  & 0.2  \\
$i$ &  7.7  & 0.5  &  -4.8  & 0.2  \\
$i2$ &  10   & 0.5  &  -6    & 0.2 \\
$z$ &  -1.6 & 0.1  &  2.5   & 0.1  \\
\end{tabular}
\label{tab:skycor}
\end{table}

Using these new $\hat{s}^{\prime}$ sky values, 
we can correct the aperture fluxes from \cite{Betoule13},
using their standard aperture area:
\begin{equation}
  N_{APER}^{effective} = \pi \left[ 7.5 \times \langle \sigma_{IQ} \rangle \right ]^2
\label{eq:area_aper}
\end{equation}
where the effective $\sigma_{IQ}$ used is also given by the \cite{Betoule13} downloadable catalogs.

Since the sky values are fitted in units of the PSF fluxes, we instead
add this correction to PSF fluxes. These two methods are perfectly
equivalent.  The new zero point averaging scheme therefore becomes :
\begin{equation}
  zp = \left\langle m_{APER} + 2.5 \log_{10} \left( \hat{f}_{PSF} +  N_{APER}^{effective} \hat{s}^{\prime} \right) \right\rangle
  \label{eq:zp_def_sky_corrected}
\end{equation}
where $ N_{APER}^{effective}$ is defined in Eq.~\ref{eq:area_aper}.
Using these corrections, we are able to eliminate the magnitude bias
of the zero point residuals, as shown in figure \ref{fig:magflat_i}:
the zero point residuals become flat over the entire range of used
magnitudes. This is good evidence that we have, on the one hand,
properly understood the origin of this bias, and, on the other hand,
properly understood the fitted sky level.

\subsection{PSF chromaticity}
\label{sec:psf_chromaticity}

\begin{figure}[h]
  \centering \subfloat[Zero point residuals as a function of color for
  real data in $g$ band, for a zero point fit using equation
  \ref{eq:zp_def_sky_corrected}.]{
    \label{fig:psf-color-data-fit}
    \includegraphics[width=\linewidth]{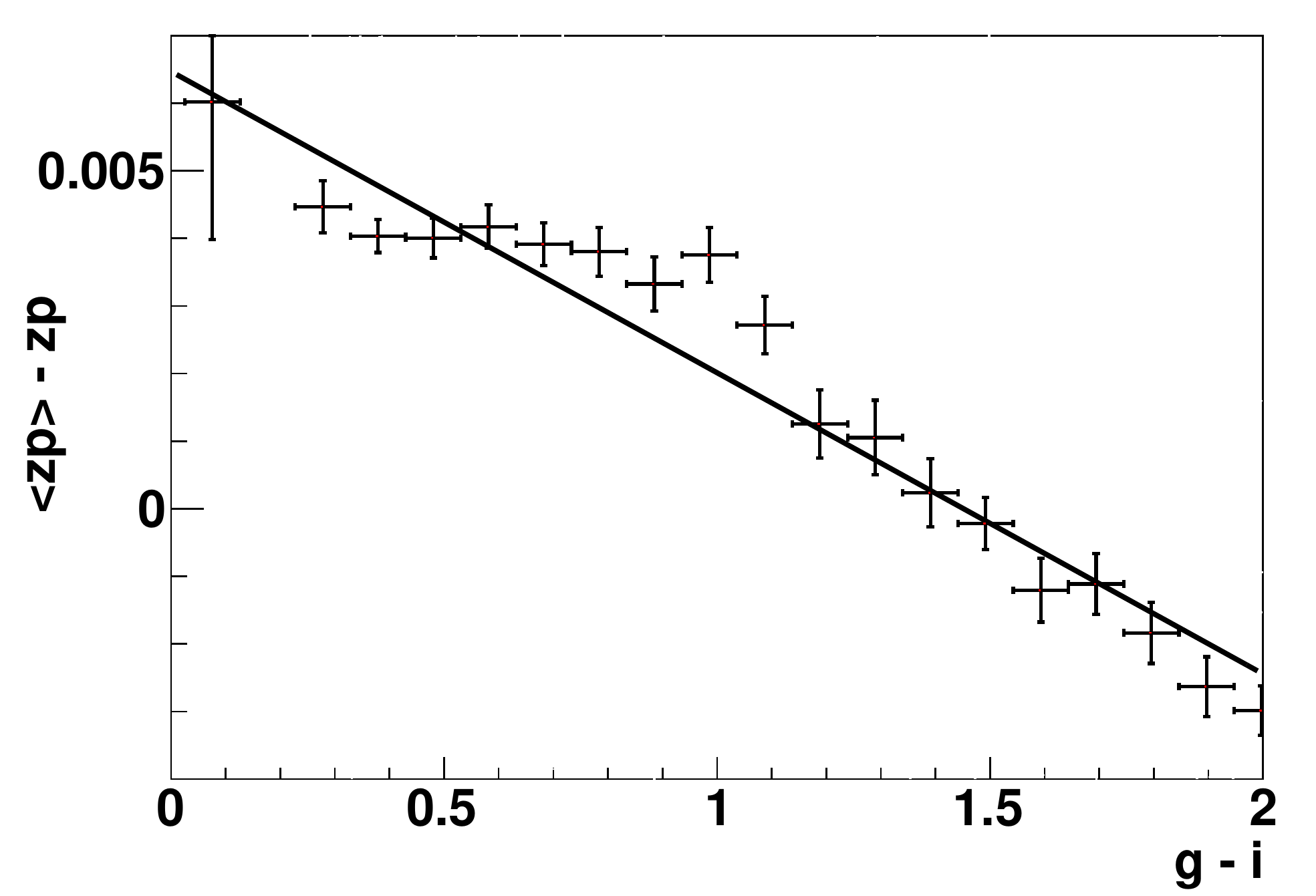}
  } \\
  \subfloat[Difference between synthetic aperture and PSF magnitudes
  as a function of the star (aperture) color, for stars from the
  \cite{Pickles98} spectroscopic library. The effective PSF bandpass
  has been adjusted for the slope to reproduce that of the above
  plot.]{
    \label{fig:psf-color-model-fit}
    \includegraphics[width=\linewidth]{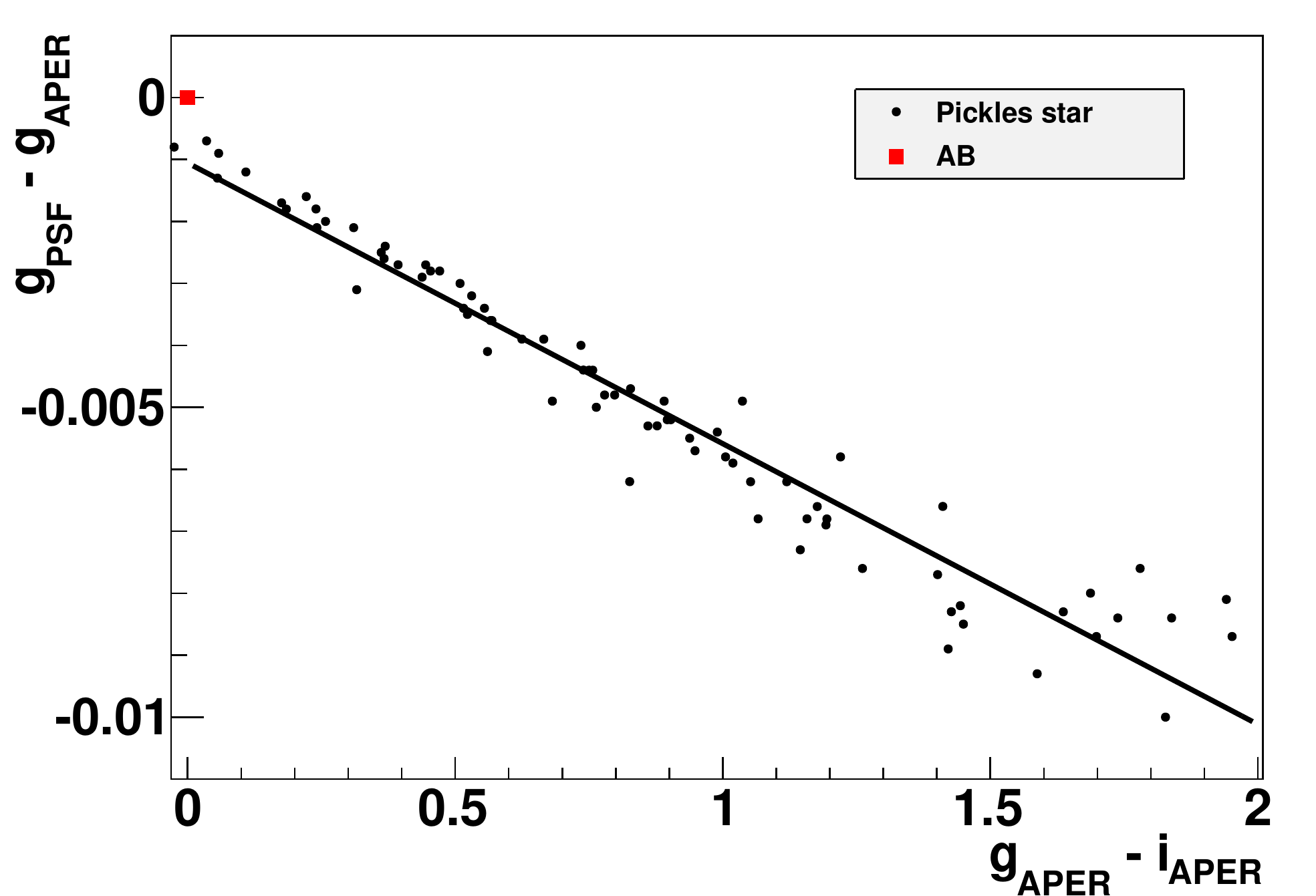}
  }
  \caption{Comparison of data with \cite{Pickles98} spectroscopic
    library in order to fit an appropriate effective filter, in $g$
    band.}
  \label{fig:psf-color-fit}
\end{figure}

In figure \ref{fig:psf-color-data-fit}, we display the values of the
zero-point residuals as a function of star color after application of
the sky level correction just discussed. A clear chromatic difference is
observed. We interpret this trend as resulting from the chromaticity
of the PSF which is not accounted for in the PSF model: blue
and red stars are measured using the same PSF model, although blue stars are
fatter than red stars (except in $z$ band where the effect is apparently reversed).
Such an effect is expected because IQ tends to improve towards red wavelengths. We call
$\alpha$ the slope of the observed relation. Its value is significant
enough that the effect must be corrected, in particular in $g$ band. We
set out to construct a natural magnitude system for PSF fluxes that
circumvents this effect. In other words, we want to be able to convert
PSF fluxes to magnitudes despite having no knowledge of the object's
color. Explicitly, we want to be able to write :

\begin{equation}
  m_{PSF} = -2.5 \log_{10} \left( \hat{f}_{PSF} \right) + zp
\end{equation}

It is clear from this requirement that PSF magnitudes will differ from
aperture magnitudes via a color term :
\begin{equation}
  m_{PSF} = m_{APER} + \alpha (c - c_{AB}) + \epsilon
  \label{eq:chrom_diff}
\end{equation}
where, for the time being, $\epsilon$ is an arbitrary offset that we
have not yet constrained. Note that in the AB magnitude system,
$c_{AB}$ is 0 by definition. We write it nonetheless to emphasize that
the relevant color term is the color difference \textit{relative to}
the standard used. Recall now the definition of magnitude for a given
spectral energy density $\phi (\lambda)$, given in
\cite{Fukugita96}. For the aperture magnitudes given in the
calibration catalog it is :
\begin{equation}
  m_{APER} = -2.5 \log_{10} \left( \frac{\int \phi(\lambda) T(\lambda) \lambda d \lambda}{\int \phi_{AB}(\lambda) T(\lambda) \lambda d \lambda} \right)
  \label{eq:APER_SED}
\end{equation}
The transmission function $T(\lambda)$ used is that derived in
\cite{Betoule13}. We choose to construct our own natural magnitude
system for PSF fluxes using an additional effective filter
$C(\lambda)$, because ignoring the PSF dependence on wavelength favours
red wavelengths over blue wavelengths (except for $z$ band where the effect
goes the other way). In other words, for a given spectral energy density
$\phi (\lambda)$, the PSF magnitude is given by :
\begin{equation}
  m_{PSF} = -2.5 \log_{10} \left( \frac{\int \phi(\lambda) T(\lambda) C(\lambda) \lambda d \lambda}{\int \phi_{AB}(\lambda) T(\lambda) C(\lambda) \lambda d \lambda} \right)
  \label{eq:PSF_SED}
\end{equation}

This effective filter is chosen in such a way that it emulates the
chromatic discrepancy between PSF and aperture magnitudes described in
equation \ref{eq:chrom_diff}. We chose to parametrize this extra-filter as a linear function of wavelength. In effect, we require that for a given
spectroscopic library of stars (here we used the \citealt{Pickles98} library), the difference between PSF and
aperture magnitudes described in equations \ref{eq:APER_SED} and
\ref{eq:PSF_SED} has the same chromatic dependency as that described
in equation \ref{eq:chrom_diff}. In figure
\ref{fig:psf-color-model-fit}, we see that the constructed linear
filter can indeed produce the required chromatic dependency. It is
clear from the definitions of $m_{PSF}$ and $m_{APER}$ that the two
yield the same value for $\phi_{AB}$. It is also clear from figure
\ref{fig:psf-color-model-fit} that the discrepancy between the two
magnitude systems is not 0 for a typical star whose color is the same
as that of AB, because of the peculiarity of the AB spectral energy
density. This constrains the free offset term $\epsilon$ in equation
\ref{eq:chrom_diff} in that it must account for this. In other words,
to convert the magnitude of a star from the aperture system to the PSF
one, in addition to a color correction term $\alpha (c - c_{AB})$, we
must also apply an offset $\epsilon$ which corresponds to the
magnitude discrepancy between the 2 systems at the color of AB. The
values obtained for $\alpha$ and $\epsilon$ are presented in table
\ref{tab:PSFchrom}, with a description of the constructed linear
filter used to obtain them.

Finally, this means that in fitting a zero point by comparing PSF
fluxes to aperture magnitudes we must take care to add a color
correction term and equation \ref{eq:zp_def_sky_corrected} becomes :

\begin{equation}
  zp = \left\langle m_{APER} + 2.5 \log_{10} \left( \hat{f}_{PSF} + \hat{s}^{\prime} \right) + \alpha (c - c_{AB}) + \epsilon \right\rangle
  \label{eq:zp_def_sky_color_corrected}
\end{equation}

\begin{table}[ht]
\caption{Color terms and offsets between PSF and aperture natural magnitude systems in each band.}
\label{tab:PSFchrom}
\centering
\begin{tabular}{ r |  r @{$\pm$} r r @{$\pm$}r r |}
Band & \multicolumn{2}{c}{$\alpha \times 10^{3}$} & \multicolumn{2}{c}{$\epsilon \times 10^{3}$} & $\lambda_0$ (\AA) \\ \hline
g &  -4.7  & 0.2 & -1.1  &  0.13 & 19227  \\
r &  -0.7  & 0.2 & 0.051 &  0.013 & 57500 \\
i &  -2.4  & 0.2 & 0.82  &  0.12 & 8206 \\
i2 &  -3.1  & 0.2 & 0.98  &  0.14 & 7389 \\
z &  0.7   & 0.2 & -0.25 &  0.03 & -48409 \\ 
\hline
\end{tabular}
\tablefoot{$\alpha$ and $\epsilon$ are defined by Eq.~\ref{eq:chrom_diff}.
The $\lambda_0$ parameter describes the corresponding additional effective filter of equation \ref{eq:PSF_SED} such that $C(\lambda) = \lambda + \lambda_0$.}
\end{table}

Switching spectroscopic libraries changes the value of 
$\epsilon$ by about $10^{-4}$ magnitude or less depending on the
band. There is therefore no significant systematic error associated
with this magnitude system transformation.

 Finally, we consider the implications that PSF chromaticity
  can, in part, be due to atmospheric effects, and that the resulting
  color term might then vary with IQ. We have fitted the slope of
  figure \ref{fig:psf-color-data-fit} separately for IQs below and
  above the median IQ, and have found extremely compatible values.  We
  hence conclude that a single color term can effectively describe the
  chromatic effects independently of IQ.

\section{SN photometry}
\label{sec:sn-photometry}
\subsection{Performance on SNLS supernovae}

The SNLS observing strategy calls for multiple images per night. By
assuming that the supernova flux does not evolve significantly during
the night, we can fit a new lightcurve where all fluxes in the same
night have been averaged. Namely, we minimize 
\begin{equation}
\chi^2 = (F_i -AF_n)^T W_i (F_i -AF_n) \label{eq:chi2_night}
\end{equation}
over the vector of fluxes per night $F_n$, where $F_i$ 
if the vector of measured fluxes over individual images, 
$W_i$ the inverse of their covariance matrix, and $A$ is 
a rectangular matrix filled with 1's at positions that assign images
to nights, and 0's elsewhere. The $\chi^2$ of such a fit yields a good
quality test of the photometry used, because it measures
the compatibility of fluxes measured over the same night. 
Note that this fit is weighed
without taking into account the shot noise of the SN itself. Because
the shot noise is estimated using the signal itself, such a weighing
scheme would be biasing, because negative
fluctuations would receive larger weights than positive ones \citep[see e.g.][and references therein]{Humphrey09}.

\begin{figure}[h]
  \centering
  \includegraphics[width=\linewidth]{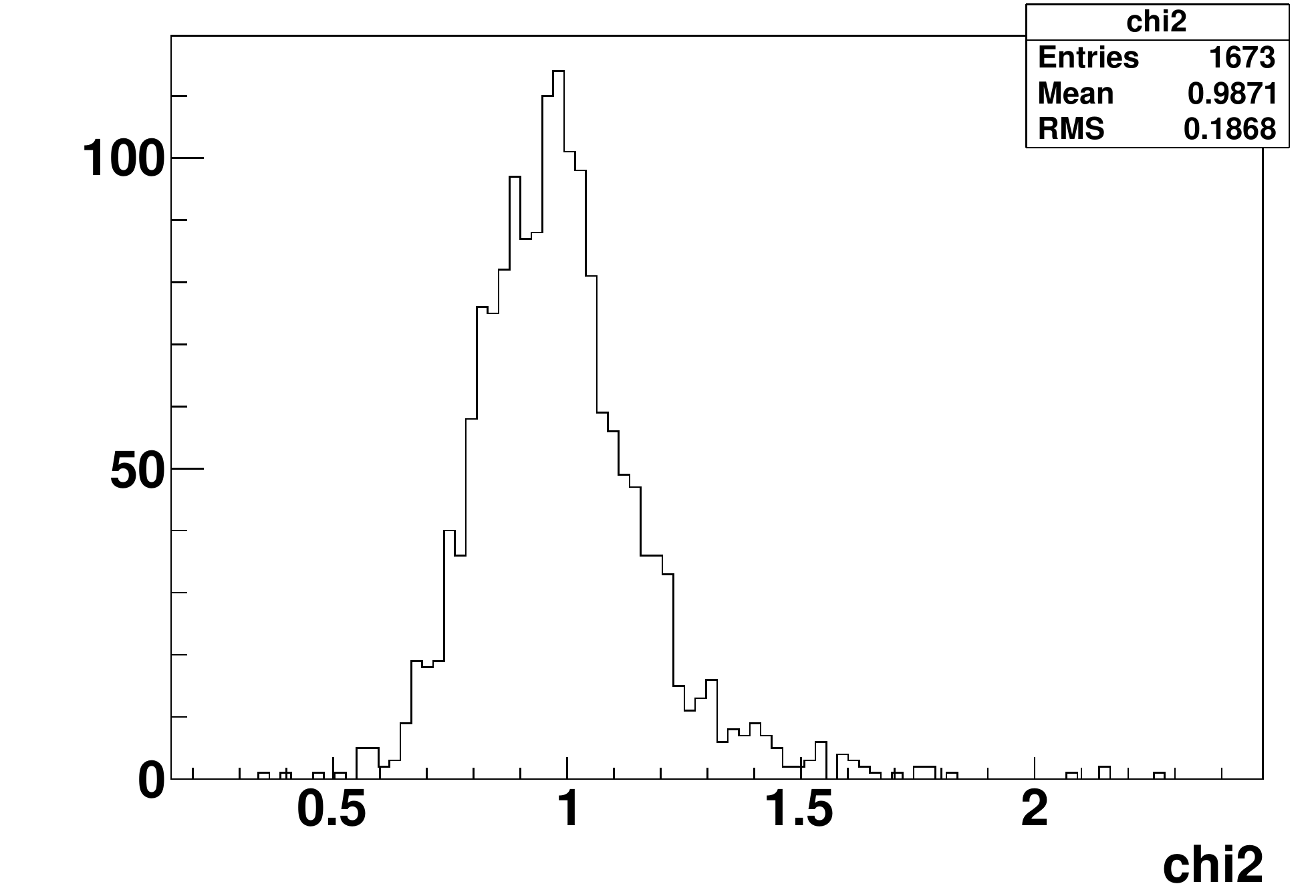}
  \caption{Distribution of the  $\chi^2/N_{\mathrm{dof}}$ of night fits (Eq.~\ref{eq:chi2_night}) of real SNe.}
  \label{fig:chi2_distri}
\end{figure}

\begin{figure}[h]
  \centering
  \includegraphics[width=\linewidth]{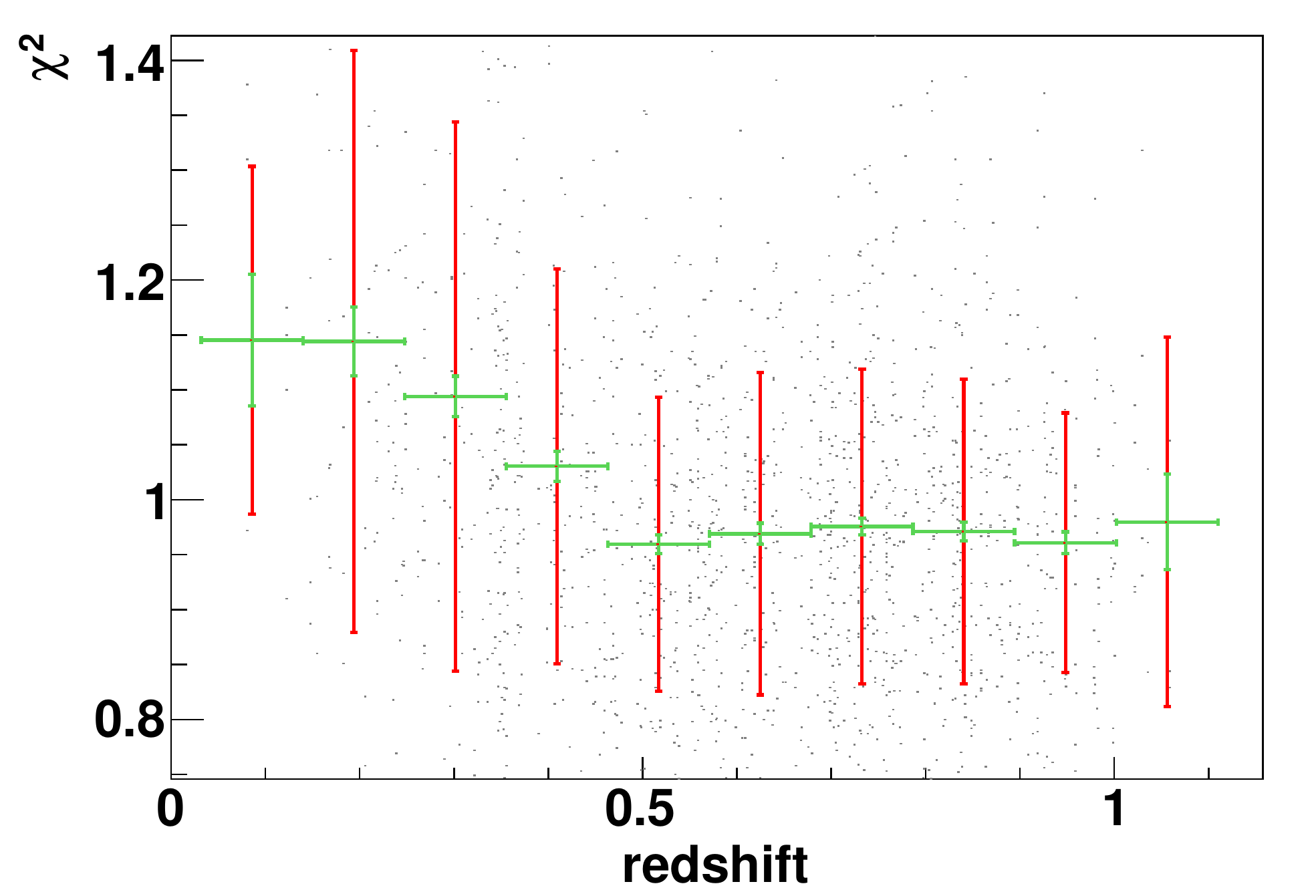}
  \caption{Evolution of  $\chi^2/N_{\mathrm{dof}}$ of night fits  (Eq.~\ref{eq:chi2_night}) of real SNe with their redshift.}
  \label{fig:chi2_VS_z}
\end{figure}

In figure \ref{fig:chi2_distri} we provide the $\chi^2/N_{\mathrm{dof}}$ distribution of
these fits and the average value is very close to 1. This is
particularly true for high redshift SN, where uncertainty terms scaling
with flux are negligible. Indeed, we see in figure \ref{fig:chi2_VS_z}
a clear increase of the
 $\chi^2/N_{\mathrm{dof}}$ for low redshift SN, where this noise is no longer completely
negligible.
Note that the
dispersion is as expected for a $\chi^2$ distribution, given the number
of degrees of freedom for each fit.

\subsection{Band-to-band position transformations}
\label{sec:coordinate_transfer}
A significant difference between SN and tertiary star photometry comes
from the fact that the SN position is not fitted in $g$ and $z$ bands due
to the expected low S/N. Instead, we transfer the fitted
position from the $r$ and/or $i$ bands. To transfer the SN position
from, for example, $r$ to $g$, we evaluate the position of stars in both
astrometric catalogs at the epoch of the reference image in $g$ band
using the fitted proper motions, fit the geometric transformation that
maps $r$ positions to the ones in $g$, and apply this transformation to the SN position
in $r$ band. We then fit the SN in $g$ at this fixed position.

\begin{figure}[h]
  \centering
  \includegraphics[width=\linewidth]{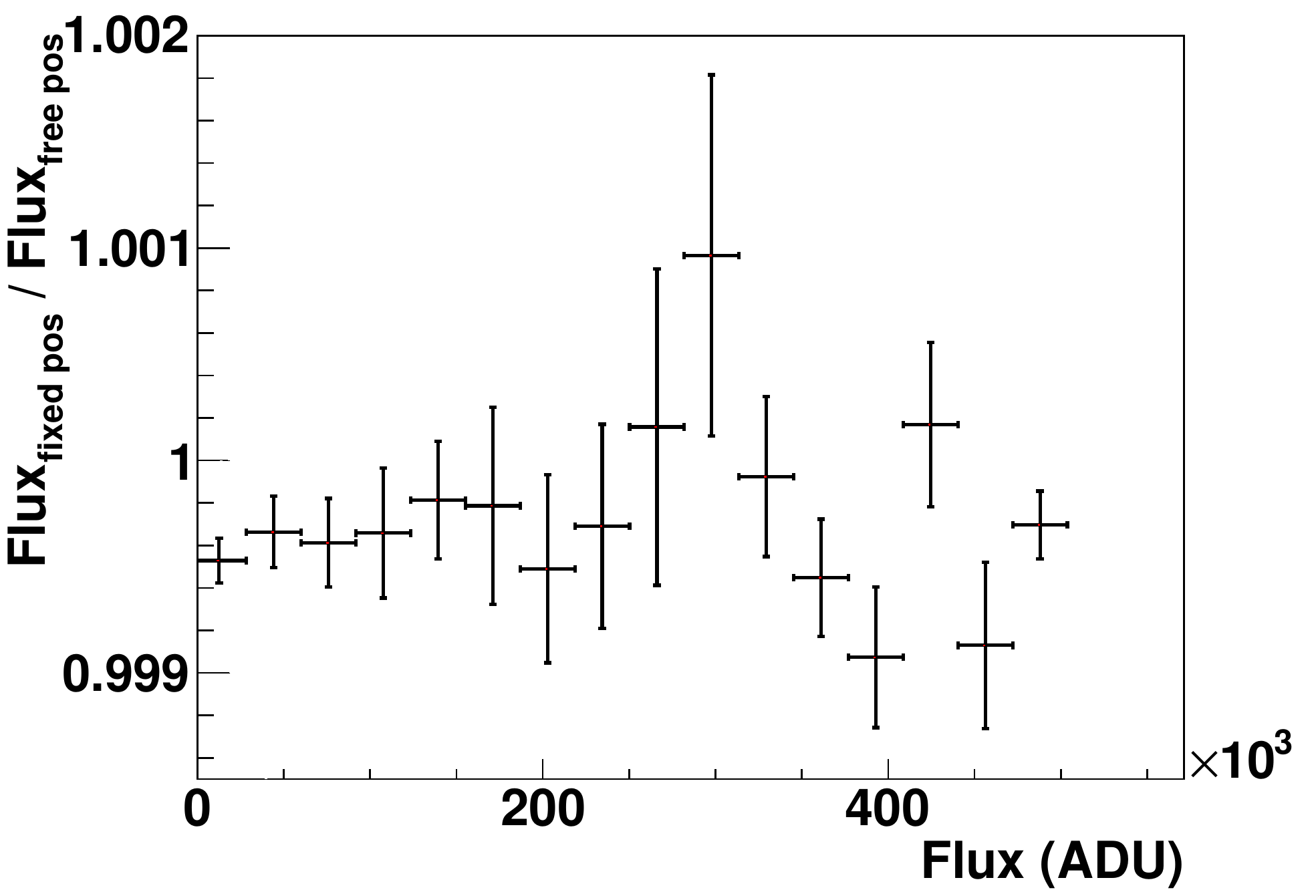}
  \caption{Ratio of the fitted flux at fixed position to
    the fitted flux with free position as a function of flux. We see
    no clear flux dependence in the flux underestimation resulting
    from the transformation.}
  \label{fig:transfoflux}
\end{figure}

\begin{figure}[h]
  \centering
  \includegraphics[width=\linewidth]{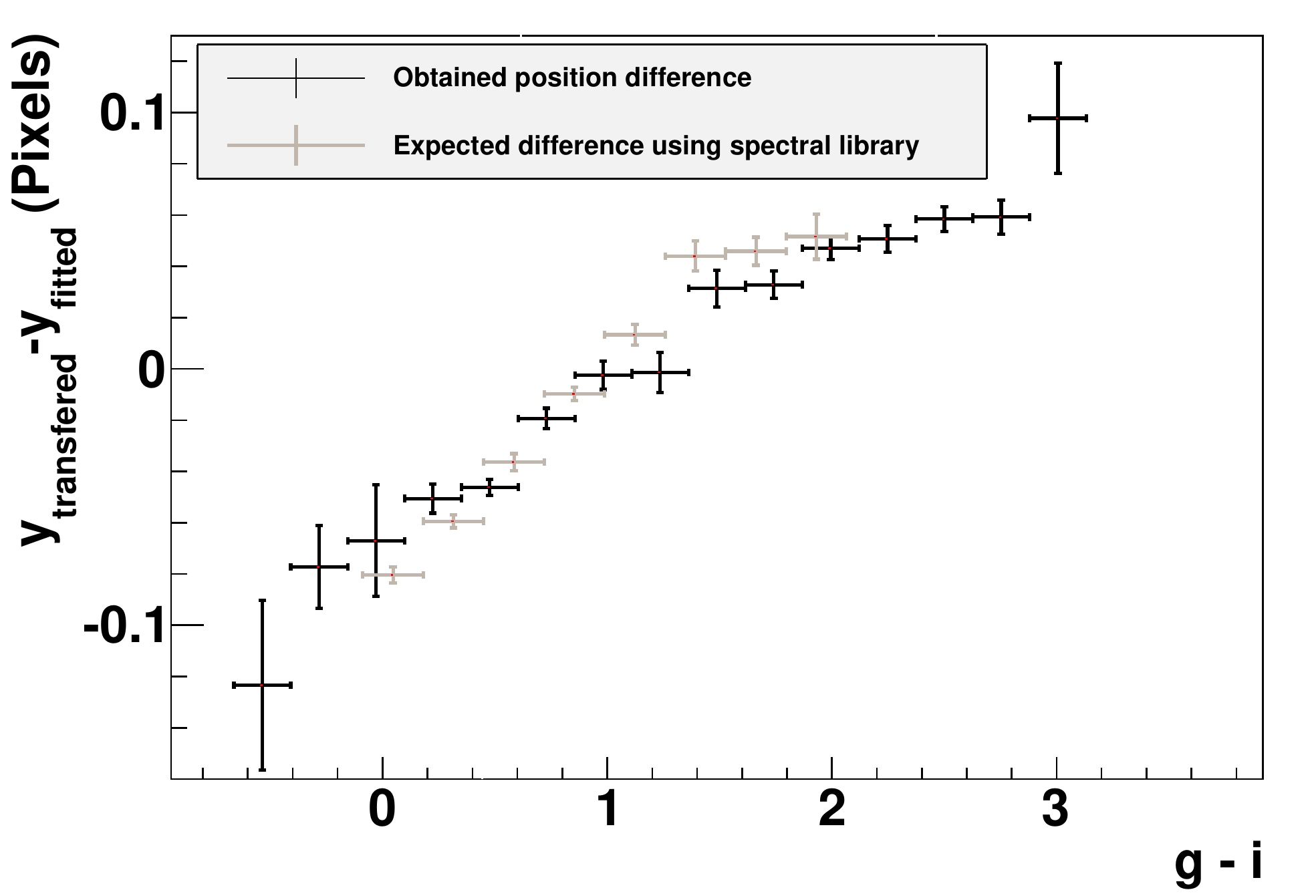}
  \caption{Difference between the fitted position in $g$
    band and the transformed position from $r$ band as a function of
    color. This is done for all stars in the D1 field. We also plot
    the expected difference obtained by computing the atmopheric
    refraction shifts for the stars in \cite{Pickles98}.}
  \label{fig:coloryshift}
\end{figure}

To ensure that this does not lead to significant differences between
SN and tertiary star photometry, we do the same for a sizable number
of tertiary stars (more precisely : all tertiary stars in $g$ band in
the D1 field). In other words, we compute light curves for these stars
at fixed positions using their fitted position in r band, after
transferring it in the same way we do for SNe. We then compare the
light curves obtained with those fitted in the regular way,
i.e. fitting both fluxes and position. In figure
\ref{fig:transfoflux}, we find that the transformation incurs a flux
underestimation of about $4 \times 10^{-4}$ independently of the flux
of the object considered. In comparing the difference between the
fitted position in $g$ band and the transferred position from r band,
it is clear that the difference is dominated by the y coordinate
term. In figure \ref{fig:coloryshift}, we see a clear trend between
the discrepancy in y and the color of the star, pointing to a
refraction effect.

Note that, as discussed previously
(\S~\ref{sec:atmospheric-refraction}), when fitting astrometry within
an image series (i.e. images in the same band) we are sensitive to the
scatter of the refraction displacements about the average
position. Here, however, we are transferring a position from one band
to another, and we are therefore concerned with the difference in the
average displacement between the 2 bands. Because for the SNLS we have
that $\mathrm{E}[\tan z \cos \eta]$ is much greater than $\mathrm{E}[\tan z \sin \eta]$,
we can understand why, recalling equations \ref{eq:deltax} and
\ref{eq:deltay}, the effect is much greater along y than along x. In
table \ref{tab:trig}, we provide the expectation values and RMS of
$\tan z \cos \eta$ and $\tan z \sin \eta$ across all fields and
filters. We find that we do not expect the effect to be much more
significant in other fields, and other bands are less affected. Using
the \cite{Pickles98} library, and computing the atmopheric refraction
shifts for a standard Mauna Kea air column at $\tan z \sin \eta =
0.48$, we are able to reproduce the slope displayed in figure
\ref{fig:coloryshift}.

The effect is most important when transporting coordinates into $g$
band, because the variation of the refractive index of air decreases
with wavelength. In order to assess the effect on supernovae, we
propagate the average $g-i$ color of supernovae as a function of
redshift $z$ ($g-i \simeq 3.6\times z -0.75$) through the relation
displayed in figure \ref{fig:coloryshift}, and input the found
displacement into expression \ref{eq:deltaF}. We find a relative flux
bias that varies from $0.5\ 10^{-3}$ at $z=0.2$ (lowest SNLS
redshifts) to $\sim 0$ at $z=0.6$ (beyond which $g$ is no longer used
to estimate distances). This refraction-induced bias is hence
negligible.

\begin{table}[ht] 
\caption{Average and standard deviation of $\tan z \cos \eta$ and $\tan z \sin \eta$ for each field, across all bands.}
\centering
\begin{tabular}{ r | r r | r r|}
 & \multicolumn{2}{|c|}{$\tan z \cos \eta$} & \multicolumn{2}{|c|}{$\tan z \sin \eta$} \\
Field & av. & rms & av. & rms \\
\hline
D1 & 0.031 & 0.46 & 0.48 & 0.040 \\
D2 & -0.049 & 0.49 & 0.36 & 0.043 \\
D3 & 0.110 & 0.53 & -0.59 & 0.056 \\
D4 & -0.075 & 0.38 & 0.79 & 0.031 \\
\hline
\end{tabular}
\tablefoot{The values do not vary significantly from band to band, so we do not further subdivide the table into bands.}
\label{tab:trig}
\end{table}

\section{Does the PSF size evolve with brightness?}
\label{sec:brighter_fatter}
The photometry methods we have been discussing so far assume 
that the PSF shape does not vary with flux at fixed color. If it does,
the flux ratios of supernovae to field stars become biased.
We now study how firmly this hypothesis is confirmed by measurements. 

\begin{figure}[hh]
  \centering
  \includegraphics[width=\linewidth]{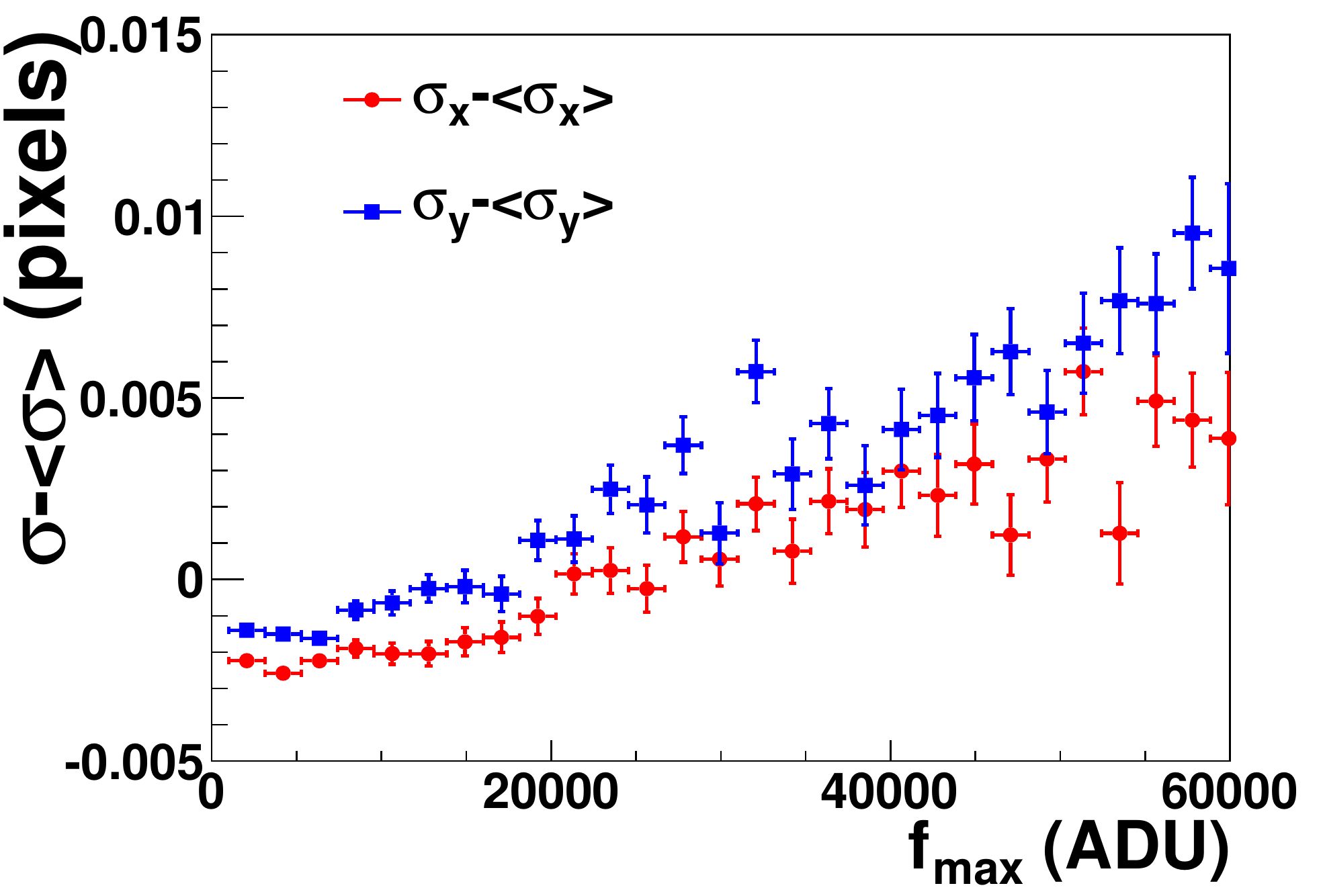}
  \caption{Difference of second moments of stars to the average (over
    1 CCD), as a function of peak flux. This is measured on a small
    set of $r$-band frames taken at low Galactic latitude, which have
    $<\sigma_{IQ}> \simeq 1.8$~pixel. From vanishing to almost saturating stars, 
    the apparent size of stars seem to vary essentially linearly with peak flux
    by about 0.5\%.
    \label{fig:fmax-sx-sy}}
\end{figure}

In figure \ref{fig:fmax-sx-sy}, we show that the apparent size of
stars, defined from their Gaussian-weighted second moments (implicitly
defined by equation \ref{eq:gaussian-moments}) tend to grow linearly
with the peak flux ($f_{max}$) of this star. This ``brighter-fatter'' relation 
is however shallow: sizes change by about 0.008 pixel over the whole brightness 
range, i.e. less than 0.5\% for this sample.

To first order, the relative flux biases from PSF photometry are
exactly the ratio of the assumed PSF size to the true size. One can
check that statement for both Gaussians and Moffat PSFs. In what
follows, we discuss the relative size change
$\sigma/\langle\sigma\rangle-1$ of a given star with respect to a
local average $\langle\sigma\rangle$, because this describes the
expected size of PSF flux biases from a wrong assumed PSF size.

\begin{figure}[hh]
  \centering
  \includegraphics[width=\linewidth]{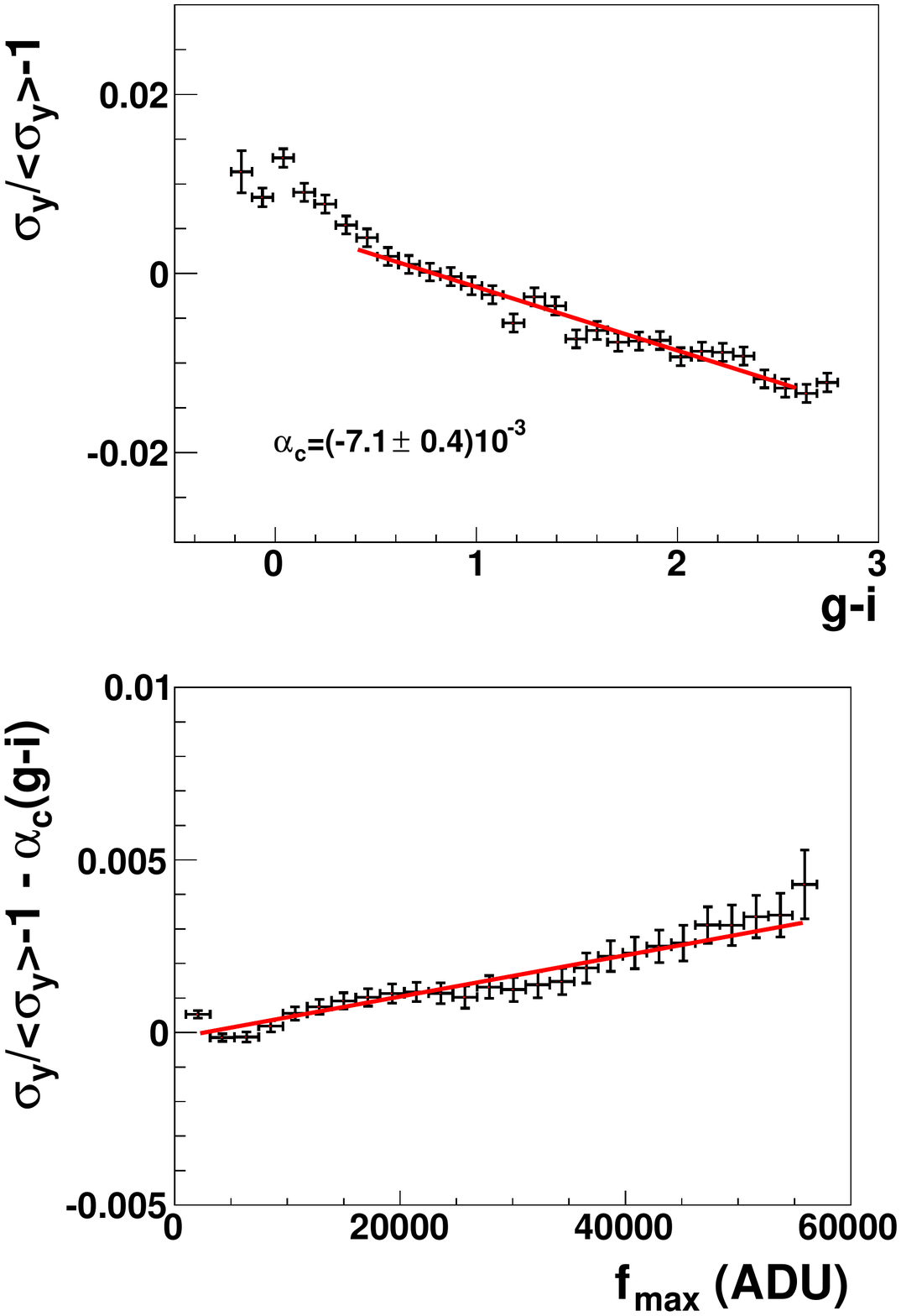}
  \caption{Relative difference of second moments of stars to the average (over
    1 CCD), as a function of $g-i$ color (top), for faint stars, and
    as a function of peak flux (bottom), correcting for the size-color relation
    Both plots display the science data, in $g$ band. Values for the slopes 
    in all bands are provided in
    table \ref{tab:color-fmax-slope} 
    \label{fig:color-fmax-slope-ycc}}
\end{figure}

We know that, due at least to atmospheric effects and electron diffusion 
in the sensors, the ``size''
of stars (in a given band) depends on their color. Since in a given
sample of measurements, flux and colors are usually related,
we first measure this color vs size dependence by restricting the analysis 
to the low-flux regime. We then measure the actual rise of
apparent size with peak flux after accounting for the size-color relation. 
Both relations are illustrated
in figure \ref{fig:color-fmax-slope-ycc} for the $g$ band and $y$ direction, 
and the fitted slopes in all bands for both correlations and both coordinates
are provided in table \ref{tab:color-fmax-slope}. We note that 
the size-color relations are broadly compatible with the color terms
of PSF vs aperture magnitudes of table \ref{tab:PSFchrom}.
We also note that we do observe a color-independent brighter-fatter relations, which are slightly
but consistently steeper along $y$ than along $x$ and similar across bands.
Accounting for the
size-color relation is mandatory in $g$ and $r$ bands, but does not
significantly change the results in redder bands. Note that requiring
that e.g. $g-i$ is measured introduces a relation between $f_{max}$ in
$g$ band and color because bright stars in $g$ tend to
saturate in $i$ band if they are red. 

\begin{table}[h]
\centering
\caption{Slopes of size vs color and size vs $f_{max}$ relations.}
\begin{tabular}{ c | c c | c c |}
 &  \multicolumn{2}{c|}{color slope ($\times 10^{3}$)} &  \multicolumn{2}{c|}{$f_{max}$ slope ($\times 10^{7}$)} \\
\cline{2-5}
Band   & $x$ & $y$ & $x$ & $y$ \\
\hline 

$g$  & -7.7 $\pm$ 0.4 & -6.6 $\pm$ 0.5 & 0.44 $\pm$ 0.04 & 0.60 $\pm$ 0.04 \\
$r$  & -2.1 $\pm$ 0.3 & -2.1 $\pm$ 0.5 & 0.52 $\pm$ 0.03 & 0.54 $\pm$ 0.03 \\
$i$  & -2.2 $\pm$ 0.3 & -2.9 $\pm$ 0.5 & 0.34 $\pm$ 0.02 & 0.42 $\pm$ 0.03 \\
$i2$ & -3.1 $\pm$ 0.4 & -3.5 $\pm$ 0.5 & 0.43 $\pm$ 0.04 & 0.50 $\pm$ 0.04 \\
$z$  &  0.1 $\pm$ 0.3 & -0.1 $\pm$ 0.5 & 0.44 $\pm$ 0.03 & 0.48 $\pm$ 0.03 \\
\hline
\end{tabular}
\tablefoot{Slopes of fitted straight lines of figure \ref{fig:color-fmax-slope-ycc}
for all five bands along $x$ and $y$ directions. Color slopes
are broadly compatible along $x$ and $y$, and also broadly compatible
with the ones provided in table \ref{tab:PSFchrom}. The color-corrected $f_{max}$ slopes
are all slighly larger along $y$ than along $x$, and do not vary 
significantly across bands.
Uncertainties are to be regarded with some caution,
because the considered effects are small and fitted values of the slopes depend
on fitting intervals.\label{tab:color-fmax-slope} 
}
\end{table}

Because calibration stars are significantly brighter than supernovae,
their PSF flux ratio measured using the same PSF are biased due to
this PSF size change.  However, the calibration stars have a peak flux
at most of 20000 ADU, and on average of about 5000 ADU. So, given the slopes
in table \ref{tab:color-fmax-slope}, the relative flux biases between
supernovae and tertiaries are around  $3 \times 10^{-4}$, and can hence
be ignored. 

Instead of these a posteriori arguments, it would be more appropriate
to incorporate the change of PSF size with flux into the PSF model. It
would have been our policy had we realised early enough the
existence of the effect. Indeed, we discovered its existence while
analyzing test bench data from other sensors\footnote{We are grateful to P. Doherty (Harvard University) for providing us with test data of a candidate 
sensor for LSST.}. 
In both 
instances, we find that the size rise is faster by about 20\% along
$y$ (CCD columns) than along $x$ (CCD rows). We do not find evidence
of a significant variation from band to band. Because the brighter-fatter effect seems linear with flux on two different sensors, it is tempting to attribute it to a physical effect within the sensor, rather than some non-linearity of the electronic chain. One possible cause might be that 
charges stored in the CCD induced by bright objects repel forthcoming 
charges, thus causing the broadening.

\section{Practical implementation of the simultaneous photometry algorithms}
\label{sec:implemention}
The cores of both codes, written in C++, are Newton-Raphson minimizers
which require computation of both the gradient and the Hessian of the
$\chi^2$. We use analytical derivatives for all parameters,
and the Hessian computation dominates the CPU budget. The algorithms 
require an input position for the source (or the
source list), and the first minimization step is carried out at the
fixed input position in order to derive fluxes, which are required to
evaluate the derivatives with respect to position. Then, the position
is released and the required fit is carried out, which typically
converges in 5 iterations. Because it fits larger stamps, the RSP
algorithm is 2 to 3 times slower than the DSP algorithm. The latter
fits a supernova lightcurve with its underlying galaxy on $\sim$ 700
images in about 2500 s on typical $\sim$2 GHz recent computers. On the
same image set, fitting lightcurves of  $\sim$100 tertiaries (without
an underlying galaxy) takes about 2 hours. Neither of the codes has 
been aggressively optimised.

\section{Summary and conclusions}
\label{sec:conclusion}

We have presented two photometric methods to measure light curve of
supernovae in the framework of the SNLS. Both methods have been tested
using realistic simulations which consist in adding artificial
supernovae to real images by copying artificially dimmed real stars,
and comparing the measured and the applied dimming. The RSP method
exhibits a marginally significant bias of $\sim 2\pm 1$ mmag, which
survived numerous tests, and that we hence attribute to
statistical fluctuation (\S~\ref{sec:RSP_accuracy}). The
non-resampling method DSP appears free of biases to a similar
1 mmag accuracy (\S~\ref{sec:DSP_accuracy}).  We have also derived an
uncertainty model that accurately describes the scatter observed in
simulations by adding a systematic noise floor of $6\times 10^{-3}$ to
propagated shot noise contributions
(\S~\ref{sec:photometric_uncertainty}). In practise, the DSP method is
faster (mostly because it does not require resampling) and more firmly
founded from a statistical point of view (because it does not ignore
pixel correlations introduced by resampling) and should hence be
prefered over the RSP method. The DSP method also allows one to over-sample
the galaxy model while the RSP method does not.

We have established the methods required to accurately compare
aperture calibrated magnitudes with instrumental PSF magnitudes of the
field stars, required to attribute calibrated magnitudes to
supernovae. Two effects have to be corrected for: 
\begin{itemize}
\item systematic biases of PSF fluxes due to the wavelength dependence
of the PSF, ignored in the PSF model. This translates into effective
PSF bandpasses slightly different from aperture bandpasses. We model
the difference between these two sets of bandpasses using the color
terms between both sets of magnitudes (\S~\ref{sec:psf_chromaticity});
\item systematic biases
of the estimated sky level in aperture photometry due to the tertiary stars
being more isolated than average, and hence being less contaminated by 
light from other objects in the field than average objects. We correct
for this effect using the sky level measured from the simultaneous
PSF photometry (\S \ref{sec:sky_subtraction}). We first have to correct for colored systematic contributions
to this estimated sky level,  due to the wavelength dependence
of PSF mentionned just above.
\end{itemize}
From figures \ref{fig:zp_VS_mag_i} and \ref{fig:psf-color-fit}, we generously
attribute an uncertainty of these two operations of 1~mmag. Together with
the uncertainty of 1~mmag for the SN to tertiaries flux ratio, 
magnitudes of supernovae are affected by an uncertainty of $\sim$1.5~mmag
with respect to calibrated aperture magnitudes of tertiary stars.
This is significantly smaller than the systematic uncertainties
affecting the physical fluxes of these tertiary stars,
which are of the order of $\sim$4-5~mmag \citep[Table 23 of][]{Betoule13}.

Our photometric method requires a determination of the relative astrometry of
individual images involved in the measurements and we reach a
positional floor uncertainty of 2.4~mas per star, image and
coordinate, independent of the band
(\S~\ref{sec:astrometric-fit}). This precision allows us to detect
tiny variations of CCD row physical size
(\S~\ref{app:pixel-size-variations}).  We investigate effects of
refraction on relative astrometry within a band
(\S~\ref{sec:atmospheric-refraction}) and across bands
(\S~\ref{sec:coordinate_transfer}) and conclude that it can be ignored
in both cases. We eventually detect that bright stars tend to appear
slightly ``fatter'' than faint stars, but this effect does not
significantly affect our photometry and calibration procedure.

\begin{acknowledgements}
J. Marriner kindly accepted to read the manuscript, and we gratefully
followed most of his suggestions. This paper relies on a very large
set of high-quality images acquired by Queue Service Observing team at
CFHT. This research used the facilities of the Canadian Astronomy
Data Centre operated by the National Research Council of Canada with
the support of the Canadian Space Agency. The authors heavily rely on
the Centre de calcul de l'IN2P3 for computing and data storage.
\end{acknowledgements}

\appendix

\section{Form Factor for Displacement Errors}
\label{sec:formfactor}

To get to equation \ref{eq:deltaF} we rely on the key assumption that the PSF is nearly identical to a Gaussian for which $\sigma _x = \sigma _y = \sigma _{seeing} ~\&~ \sigma _{xy} = 0$. To account for errors induced by this assumption, we allow for an additional form factor $F$ in equation \ref{eq:deltaF} which then takes the form:

\begin{equation}
	\mathrm{E}[\hat{f}] = f \left\{ 1 - F \times \frac{(\Delta x)^2 + (\Delta y)^2}{4 \sigma ^2} \right\} \label{eq:formfactor}
\end{equation}

To find the value of $F$, we use the exact expression for the flux estimator expectation for any given PSF model which is :

\begin{equation}
  \mathrm{E}[\hat{f}] = f \frac{\mathrm{E}[\int \overbrace{PSF(x,y)}^{Data} \overbrace{PSF(x-\hat{\delta _x},y-\hat{\delta _y}}^{Fitted PSF}) dx dy]}{\int PSF^2(x,y) dx dy} \label{eq:PSFerr}
\end{equation}

We then compute equation \ref{eq:PSFerr} numerically for a range of displacements using the PSF model of the reference image. When fitting equation \ref{eq:formfactor} to the results we obtain $F = 0.788$.

\section{Pixel size variations}
\label{app:pixel-size-variations}
\begin{figure}[h]
\centering
\includegraphics[width=\linewidth]{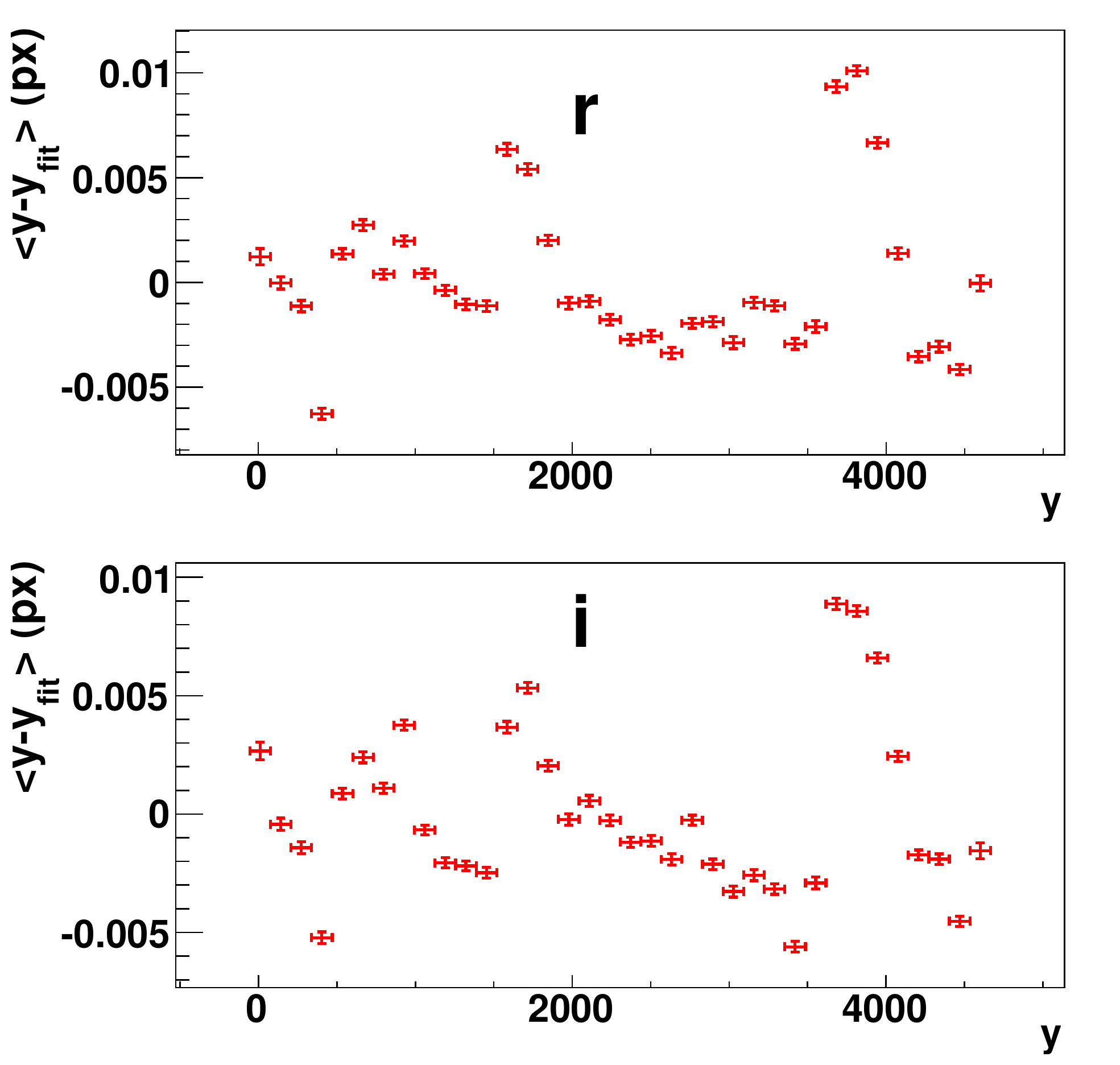}

\caption{Average astrometric residuals along $y$ (expressed in pixels)
  as a function of the $y$ coordinate of measurements, for
  measurements in CCD 12 of stars brighter than 20.5. The average runs
  over the 4 SNLS science fields, i.e. $\sim$2000 $r$ images and
  $\sim$3000 $i$ images. The shape is very similar in $r-$ and $i-$
  bands, suggesting some instrumental source. The jumps of residuals
  happen at $y$ values close to multiples of
  512.\label{fig:plotir_residuals} }
\end{figure}

\begin{figure}[h]
\centering
\includegraphics[width=\linewidth]{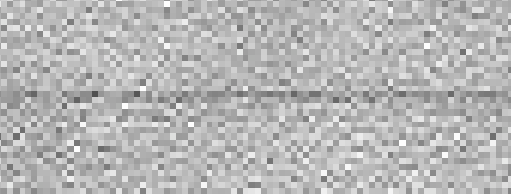}
\caption{Stamp of a flat-field image around $y=515$.
The response variation around 
one or two rows is clearly visible, corresponding to the discontinuity
of figure~\ref{fig:plotir_residuals}. 
\label{fig:flat-image}
}
\end{figure}

\begin{figure}[h]
\centering
\includegraphics[width=\linewidth]{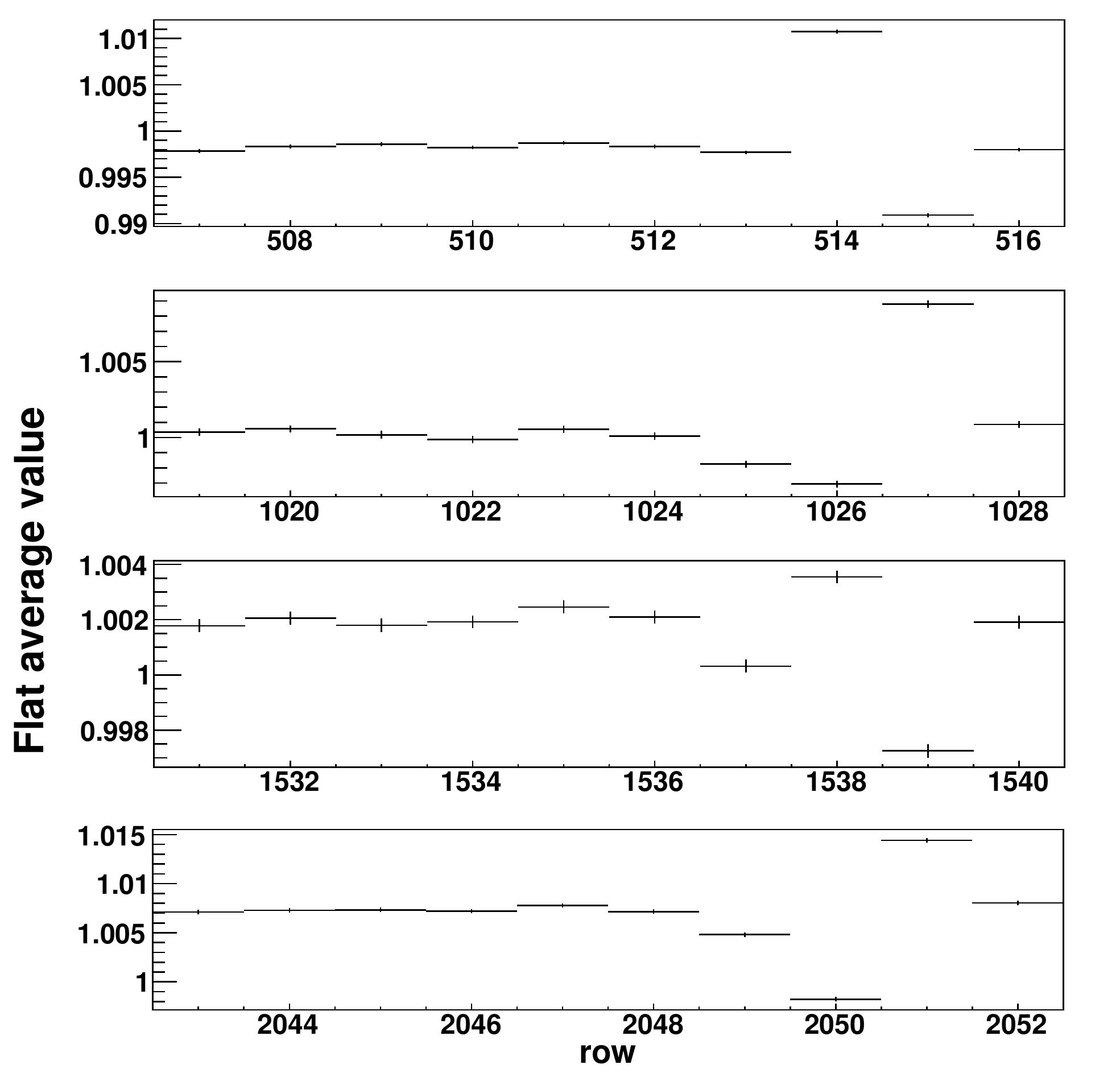}
\caption{Flat-field average value along rows as a function of $y$,
  zoomed around some of the discontinuities of figure
  \ref{fig:plotir_residuals}.  The response variation extends over a
  few rows and is typically of the order of 1\%. They are located at
  $y \simeq 3+n\times 512$.
\label{fig:flat_wiggles}
}
\end{figure}

The astrometric fit we have described in \S~\ref{sec:astrometric-fit}
assumes that positions measured in pixel units describe the physical
position in the CCD, perhaps up to some very smooth variation. This is
not necessarily realised: the physical pixel size might
vary too rapidly for the coordinate mappings to accommodate the
variation. The manufacturing of CCDs can make some rows or columns
wider or narrower than the average. Figure \ref{fig:plotir_residuals}
presents the average astrometric residual along $y$ as a function of
the $y$ position of the measurement.  It exhibits discontinuities of
the order of 0.01 pixel, observed at the same locations in both $r$
and $i$ bands, and at $y$ values roughly multiples of 512. Figure
\ref{fig:flat-image} displays a stamp of a flatfield image ($r-$band)
where the variation of flat-field response along one or two rows is
clearly visible. The flat-field average value over rows is displayed
in figure \ref{fig:flat_wiggles}, where one can spot variations of the
order of 1\% of the average flat values at the positions where we
detect astrometric residual discontinuities. It is tempting to attribute
the latter to physically wider or narrower rows, due to some tiny
misplacement of the masks during the CCD manufacturing (type E2V
CCD42-90). As we do not detect any comparable pattern along $x$, and
since the residuals along $x$ are not smaller than along $y$, we cannot
attribute a sizable fraction of the astrometric noise floor to these
small defects. We thus did not attempt do incorporate those into the
astrometric model. Similar mechanical defects with the same
consequences on astrometry were discussed about the WFPC2 camera on
the HST in \cite{Anderson99}.

\bibliographystyle{aa}
\bibliography{biblio}

\end{document}